\documentclass[a4paper,fleqn,usenatbib]{mnras}

\usepackage{graphicx}
\usepackage[percent]{overpic}
\usepackage{lipsum}
\usepackage{multirow}
\usepackage{color}
\usepackage{rotating}
\usepackage{mwe,tikz}
\usepackage{mathtools}
\usepackage{physics}
\usepackage{amsmath}
\usepackage[utf8x]{inputenc}

\newcommand{\msun}{${\rm M_{\sun}}$}

\def\ltsima{$\; \buildrel < \over \sim \;$}
\def\simlt{\lower.5ex\hbox{\ltsima}}
\def\gtsima{$\; \buildrel > \over \sim \;$}
\def\simgt{\lower.5ex\hbox{\gtsima}}
\def\kms{{\rm\,km\,s^{-1}}}
\def\pc{{\rm\,pc}}
\def\kpc{{\rm\,kpc}}

\def\msun{{\rm\,M_\odot}}


\def\deg{^\circ}

\def\Gyr{{\rm\,Gyr}}

\def\ltsima{$\; \buildrel < \over \sim \;$}
\def\gtsima{$\; \buildrel > \over \sim \;$}

\title[Finding Stellar Streams]{\texttt{STREAMFINDER} I: A New Algorithm for detecting Stellar Streams}

\author[Malhan \& Ibata]{
Khyati Malhan,$^{1}$\thanks{E-mail: khyati.malhan@astro.unistra.fr}
Rodrigo A. Ibata,$^{1}$\thanks{E-mail: rodrigo.ibata@astro.unistra.fr}
\\
$^{1}$Universit\'e de Strasbourg, CNRS, Observatoire Astronomique de Strasbourg, UMR 7550, F-67000 Strasbourg, France\\
}

\date{Accepted 2018 April 9. Received 2018 April 4; in original form 2018 February 19}

\pubyear{2018}

\begin{document}
\label{firstpage}
\pagerange{\pageref{firstpage}--\pageref{lastpage}}
\maketitle

\begin{abstract}
We have designed a powerful new algorithm to detect stellar streams in an automated and systematic way. The algorithm, which we call the \texttt{STREAMFINDER}, is well suited for finding dynamically cold and thin stream structures that may lie along any simple or complex orbits in Galactic stellar surveys containing any combination of positional and kinematic information. In the present contribution we introduce the algorithm, lay out the ideas behind it, explain the methodology adopted to detect streams and detail its workings by running it on a suite of simulations of mock Galactic survey data of similar quality to that expected from the ESA/Gaia mission. We show that our algorithm is able to detect even ultra-faint stream features lying well below previous detection limits. Tests show that our algorithm will be able to detect distant halo stream structures $>10^{\circ}$ long containing as few as $\sim 15$ members ($\Sigma_{\rm G} \sim 33.6\, {\rm mag \, arcsec^{-2}}$) in the Gaia dataset.
\end{abstract}



\section{Introduction} 
\label{Introduction}

Stellar streams around galaxies are of great importance as their orbital structures are sensitive tracers of galaxy formation history and the underlying gravitational potential \citep{EyreBinney2009, LawMajewski2010}. The number of streams in principle places a lower limit on the number of past accretion events, allowing one to quantify the number of stars that are a result of hierarchical merging events. Moreover, in the case of the Milky Way, where we can obtain a full phase-space picture, knowing the orbits of a sample of streams can shed light on the distribution function of halo accretions (and hence probably of the halo itself). Dynamical modelling of such stellar streams is a promising avenue to constrain the dark matter distribution of the Milky Way and measure the lumpiness in its distribution
\citep{Ibata2002DM_TS, Johnston2002DM_TS, Dalal2002, StreamGap_Carlberg2012, StreamGap_Erkal2016, StreamGap_Sanders2016}.

Streams that are a result of tidal disruption of low mass progenitors tend to be dynamically cold and thin and are in particular of great interest for probing the dark matter. Dynamical modelling of their well defined and simple orbital structures is one of the best ways to constrain the dark matter distribution in the Galaxy \citep{Koposov2010, Ngan2014, Bovy2016GD1Pal5}. However, the general lack of reliable tangential velocities and distance measurements of the stream stars can be consistent with multiple (degenerate) solutions (see, e.g., \citealt{Varghese2011}). Dynamical modelling of the known streams using the quality of velocity information that will soon be made available in the second data release (DR2, scheduled for April 2018) of the European Space Agency's Gaia mission \citep{Gaia2012deBruijne,GaiaDR12016} can be used in combination with distance estimates (derived from Gaia photometry or other surveys like the Canada-France Imaging Survey, \citealt{CFIS_I_2017}) to resolve this degeneracy to some extent. But, in order to significantly improve the estimates of the Galactic mass distribution and the distribution function of the halo out to large Galactic radii, where the potential is basically unconstrained by other tracers, more  stream detections are required. The present contribution aims to construct an optimised algorithm to detect stream structures.

There already exist some effective stream detection methods that have been successful in detecting the streams that we know of so far in the Milky Way. 
These include:
\begin{enumerate}
\item Matched filter: The matched filter (MF) technique \citep{Rockosi2002,Balbinot2011MF} incorporates colour-magnitude weighting of stars to find structures that belong to a specific Single Stellar Population (SSP) model. The Palomar 5 stream \citep{Odenkirchen2001}, GD-1 \citep{Grillmair2006}, Orphan \citep{Belokurov2006}, Lethe, Cocytos, and Styx \citep{Grillmair:2016ju}, and most recently the Eridanus and Palomar 15 streams \citep{Eridanus_pal15_2017} and the 11 new streams detected in the DES \citep{DES_11_streams2018} were all found with this technique. However, the drawback of this method is that it does not incorporate kinematics and its performance is expected to drop significantly if the structure possesses a significant distance gradient.
\item Detection of co-moving groups of stars: Several halo substructures were initially identified as groups of stars of similar type (e.g. RR Lyrae, Blue Horizontal Branch Stars) that are contained within a small phase-space volume.  Several streams in the Milky Way have been detected by employing this or a variation of this technique (Aquarius by \citealt{Aquarius_Williams_2011ApJ...728..102W}, Arcturus by \citealt{Arcturus_Arifyanto_2006} and the Virgo stream by \citealt{Virgo_Duffau_2006}). The drawback of this approach lies in the fact that it requires the stars to have complete kinematic information. This requirement will not be completely fulfilled in the Galactic halo (where the streams of interest for dark matter studies lie) even in Gaia DR2. 
\item Pole counts: The Pole Count technique \citep{Johnston1996}, works well for identifying substructures that are on great circle paths around the Milky Way and are of high contrast (it was useful in detecting structures like the Sagittarius stream \citealt{Ibata2002PoleCount}). This method can be further improved by supplying the algorithm the available kinematic information \citep{Mateu2017}. The method is expected to reveal only those streams that lie almost along great circular paths on the sky, and the streams on rather complex orbits can again go undetected.
\end{enumerate}

However, in light of the revolutionary dataset that Gaia will deliver, we desired to build an algorithm that is able to use as much as possible of our prior knowledge of stellar streams to maximise the detection efficiency.
In this paper we introduce the \texttt{STREAMFINDER} algorithm that we have built, explain the physical motivation behind it and demonstrate its workings by running a suite of test simulations. We find that our algorithm can detect very faint stream features in the dataset of the quality that will soon be delivered by Gaia DR2. 

This paper is arranged as follows. In Section \ref{sec:STREAMFINDER} we present the motivation and the basic idea behind the workings of our algorithm. Section \ref{sec:Orbital_Stream_models} gives a proof of concept of our method through the detection of a simplistic orbital stream model. Section \ref{sec:N_body_simulated_stream_model} presents the success of our technique by demonstrating the detection of an N-body tidal stream structure. Section \ref{sec:Multiple_Streams} exhibits the ability to detect multiple streams criss-crossing each other in a given patch of sky. In Section \ref{STREAMFINDER_tests} we detail additional criteria incorporated into the algorithm that improve the contrast of the streams. We test the detection limit of our algorithm in finding extremely faint stream structures in Section \ref{sec:Det_limit}. In Section \ref{sec:DB_model4} we study the effect of assuming a wrong Galactic mass model. Finally, in Section \ref{sec:Conclusion} we discuss the implications of our study.

\section{\texttt{STREAMFINDER}}\label{sec:STREAMFINDER}

Different surveys of the Milky Way cover different sky regions, probe different depths of the sky and deliver different combinations of phase-space measurements. We sought to develop a generic algorithm that would work with any mix of datasets containing  any combination of positions and kinematics. We also desired the algorithm to have the property of being able to handle datasets with partial sky coverage and incomplete information on some parameters, so as to make the most of the available surveys. 

Since we suspect that the most massive star streams in the Milky Way have already been discovered, we decided to design the \texttt{STREAMFINDER} algorithm to detect primarily narrow low-mass tidal streams, and we expect these faint structures to lie hidden under a dominant ``background'' of contaminants (in most cases the contaminants will actually be in the foreground).

\begin{figure}
\begin{center}
\includegraphics[width=\hsize]{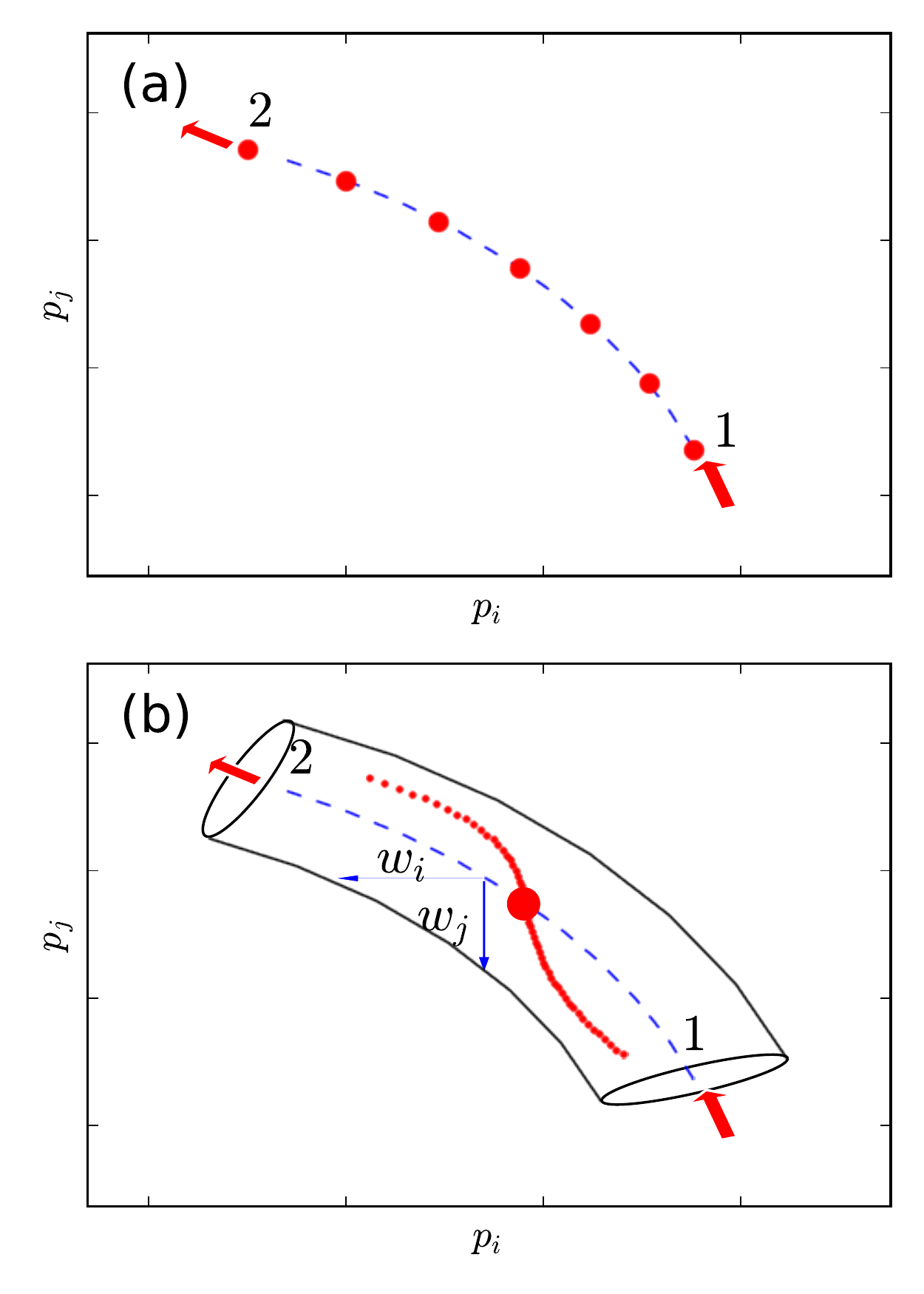}
\end{center}
\caption{The \texttt{STREAMFINDER} concept. (a) The red dots represent schematically the spatial positions along a segment of an orbit, part of a stream that we are interested in detecting. The dots labelled `1' and '2' mark, respectively, the beginning and the end of this orbit segment. The blue dashed curve represents the orbit integrated using the 6D phase-space value of stellar point `1' as initial conditions. This \textit{trial orbit} passes close to other stream members, allowing them to be associated with the structure. (b) The red dots now represent a more realistic scenario of a stellar stream where the tidal arms and the progenitor possess slightly different energies and hence lie along different orbits. Therefore, the trial orbit (blue-dashed curve) calculated using the phase-space measurement of some stream star corresponding to some (E, $L_z$) value fails to fit the entire stream structure. But if the same 6D orbit is upgraded to a 6D \textit{hyper-dimensional-tube} (black cylinder), then the stream becomes circumscribed within it.}
\label{fig:Hypertube_Diagram}
\end{figure}

\subsection{Stream Detection Concept}\label{secsub:Stream_Detection_Idea}

The tidal disruption of low-mass progenitors leads to the formation of thin and dynamically cold streams. These streams closely delineate orbits in the underlying gravitational potential of the Galaxy \citep{Dehnen2004thinorbit}. 

Consider an ideal scenario where we have a segment of an orbit (Figure \ref{fig:Hypertube_Diagram}a). The red dots represent the positions of the stars (members of a hypothetical stream that perfectly delineates this orbit) along their orbital structure in 6D phase-space. Suppose we have access to perfect 6D position and velocity values ($\mathbf{x}$, $\mathbf{v}$) for all these stream stars and that we also know the underlying gravitational potential. Then, if one integrates a \textit{trial orbit} (blue-dashed curve) using the given 6D phase-space value ($\mathbf{x_i}$, $\mathbf{v_i}$) of one of these stream stars, then this trial orbit would sew through the remaining stars in the 6D phase-space, revealing the entire stream structure.

In reality, streams do not delineate perfect orbits (Figure \ref{fig:Hypertube_Diagram}b). Stars in a tidal stream have slightly different (E, $L_z$) values, and therefore lie along slightly different orbits (see, e.g., \citealt{EyreBinney2011}). The slight differences in energies and orbital trajectories of the stream stars as they are lost from their progenitor lead to a finite structural stream width ($s$) in real space and velocity dispersion ($\sigma_v$) in velocity space.  

Our method makes use of the realisation that the members of a stream can be  contained within a 6D hyper-dimensional tube (or \textit{hypertube}) in phase-space, with width in real and velocity space similar to the size and velocity dispersion of the progenitor cluster. N-body disruption simulations show that the length of the stream, depends on the mass distribution of the progenitor, and the orbit and time of accretion onto the host galaxy. 

This suggests that a way to detect streams is to construct 6D hypertubes, with plausible phase-space width and length, and then count the number of stars that are encapsulated within them. This scenario is depicted in Figure \ref{fig:Hypertube_Diagram}b where red dots represent the stream stars, and the black cylinder is the hypertube surrounding the trial orbit (blue dashed curve).

\section{Orbital Stream models} \label{sec:Orbital_Stream_models}

We first present the algorithm applied to an idealised situation where streams follow perfect orbits. The very low contrast streams are added into a realistic mock dataset for Gaia (the Gaia Universe Model Snapshot, or GUMS, \citealt{GUMS2012}), and then we try to detect this faint stream feature from the stellar contamination using \texttt{STREAMFINDER}.

\begin{figure*}
\begin{center}
\hskip 1.0cm
\includegraphics[width=\hsize]{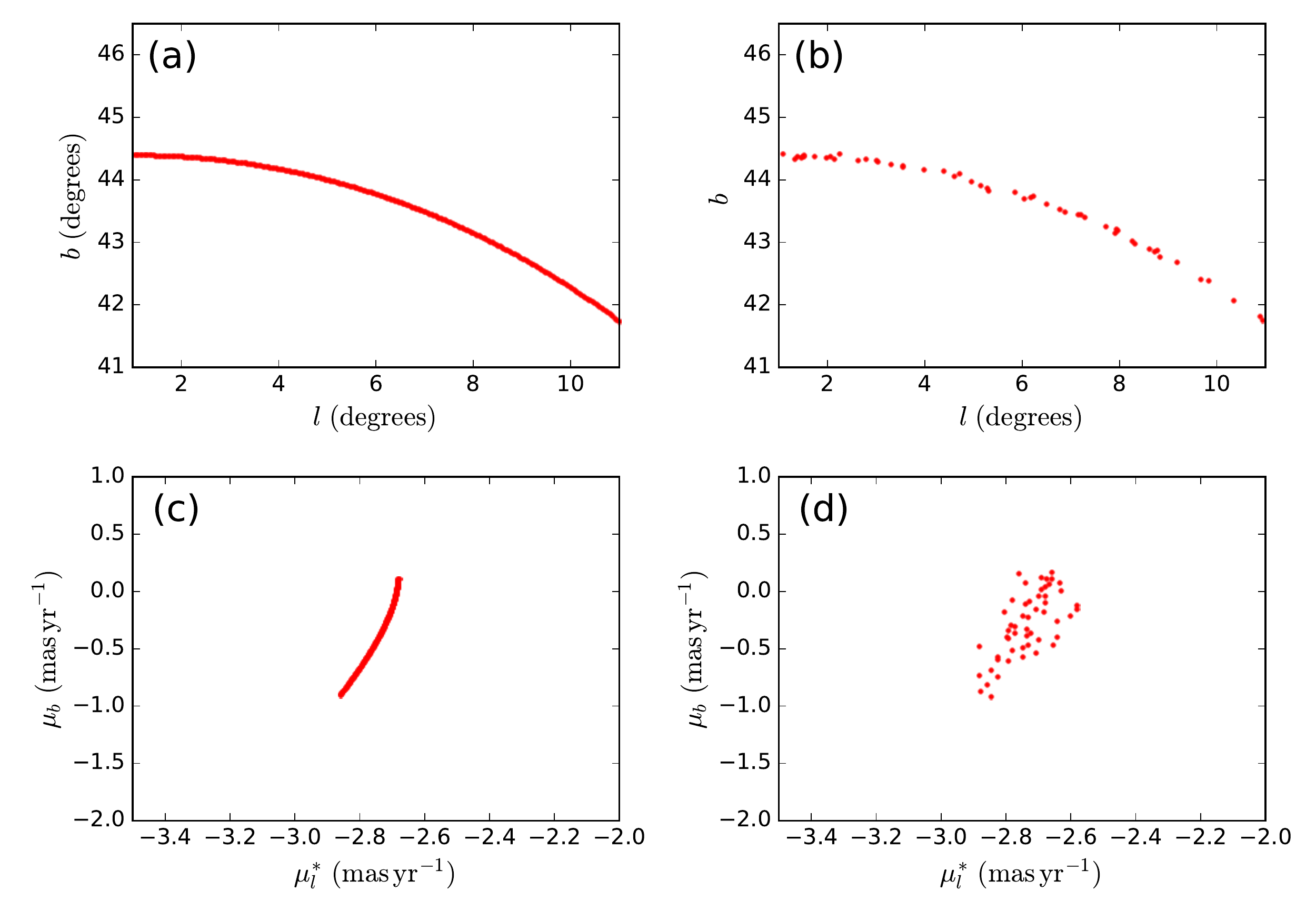}
\end{center}
\caption{Orbit stream model. The left panels show the perfect orbit model that we integrated, represented in position (a) and in proper motion (c) space. The right panels display the same orbit smeared-out to match the properties of a typical cold stream and also convolved with errors in proper motion consistent with the expected end-of-mission Gaia uncertainties. We retained only $\sim 50$ stellar points in order to obtain a low contrast structure. This structure represents the mock orbital stream model.}
\label{fig:orbit_model}
\end{figure*}

The mock stream was modelled by degrading an orbit as follows. We selected a random 6D phase-space position to give the initial conditions of the orbit. This initial condition was then integrated for $T = 0.1\Gyr$, in the realistic Galactic potential model 1 of \cite{Dehnen1998Massmodel}, to form an orbit (the value of $T$ was chosen so that the orbit appears long enough to mimic observed streams found in the SDSS). The transformation of this orbit into the heliocentric observable frame was accomplished using Sun's Galactocentric distance of $8.5\kpc$ and adopting the peculiar velocity of the Sun $\bmath{V_{\odot}} = (u_{\odot}, v_{\odot}, w_{\odot}) = (11.1, 12.24, 7.25) \kms$ \citep{Schornich2010_Sun}. The resulting  orbit model is shown in Figure \ref{fig:orbit_model}.

To give this orbit a stream-like appearance, we need to provide a structural width and a velocity dispersion. For this, we chose $s = 50\pc$ and $\sigma_{v} = 2\kms$. These values are adopted in accordance with the values of some of the currently known dynamically cold streams (\citealt{Grillmair_book_2016}, and references therein). To smear the data in phase-space, every orbital point was then convolved with a Gaussian with dispersion equivalent to these values. 

\begin{figure*}
\begin{center}
\includegraphics[width=\hsize]{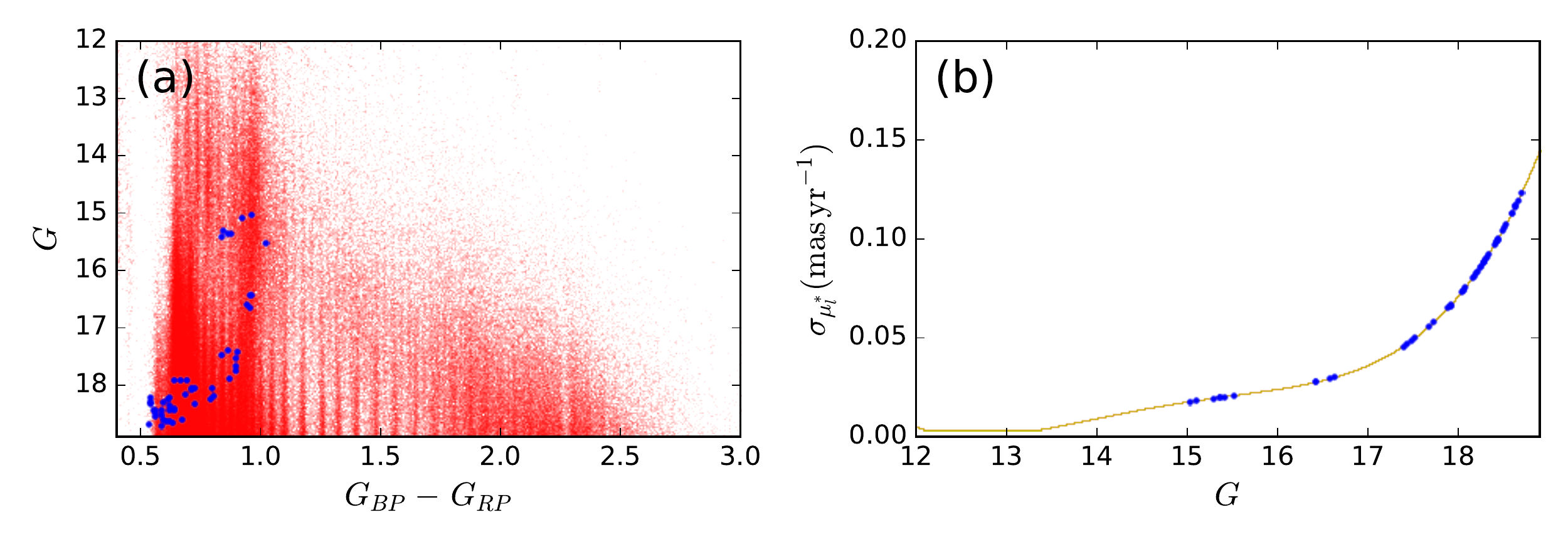}
\end{center}
\caption{Mock Gaia data. The colour-magnitude diagram of the mock Gaia dataset is shown in panel (a) where the stream stars are highlighted in blue. (The vertical stripes in the CMD are GUMS simulation artefacts). Panel (b) shows the variation of the proper motion errors in $\sigma_{\mu_l}$ as a function of G-band magnitude (orange line). The bigger blue dots represent  $\sigma_{\mu_l}$ for the stream stars. For the purpose of these tests, we assume that the uncertainties in $\sigma_{\mu_b}$ mirror those in $\sigma_{\mu_l}$.}
\label{fig:errors_on_Gaia_data_Orbit}
\end{figure*}

The stream stars were assigned ${\rm G_{BP}-G_{RP}}$ colour and $G$ magnitude in the Gaia bands, using a Padova SSP model \citep{Marigo2008Padova} of metallicity ${\rm [Fe/H]=-1.5}$ and age $10\Gyr$, appropriate for a typical halo globular cluster. A lower limit of ${\rm G_0}=19$ was chosen so as to mitigate against variations in extinction in the high latitude fields of interest for halo studies, which would otherwise cause variations in survey depth. Given the assigned magnitude, we generated an uncertainty in proper motion ($\mu_{l}$, $\mu_{b}$) according to the ``End-of-mission'' sky average\footnote{See {\tt https://www.cosmos.esa.int/web/gaia/sp-table1}} as shown in Figure \ref{fig:orbit_model}. The dependency of the proper motion errors on the G-band magnitude is shown in Figure~\ref{fig:errors_on_Gaia_data_Orbit}.

The detection limit for radial velocities in Gaia DR2 is expected to be only ${\rm G=13}$~mag, but even in the later data releases, most Gaia halo stars will not have measured radial velocities. Likewise, virtually no distant halo stars will have well-measured parallaxes with Gaia. We therefore omit both the radial velocity and distance information from our simulated streams, retaining only 4D astrometric information of the mock stream stars in the form of ($\ell, b, \mu_{l}, \mu_{b}$) along with the stellar photometry (${\rm G, G_{BP}, G_{RP}}$) and associated observational errors (Figures \ref{fig:orbit_model} and \ref{fig:errors_on_Gaia_data_Orbit}).

The GUMS data were degraded in proper motion based on their G-band magnitudes and once again we retained only 4D phase-space information of the data in the form of ($\ell, b, \mu_{l}, \mu_{b}$) along with the photometry ($G, G_{BP}, G_{RP}$) and the observational errors. 

\begin{figure*}
\begin{center}
\includegraphics[width=\hsize]{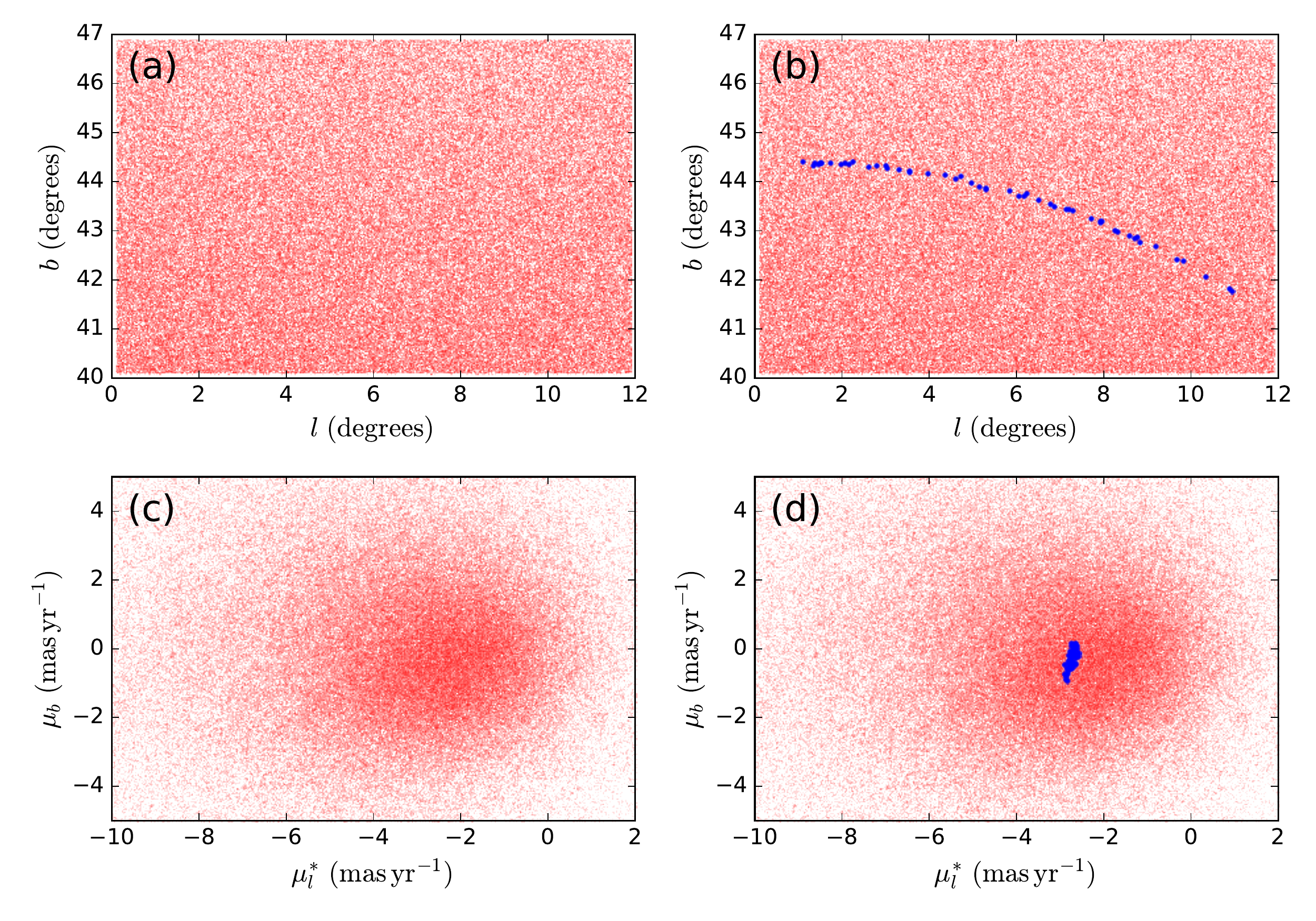}
\end{center}
\caption{Mock Gaia data. The stream orbit model (shown in Figure~\ref{fig:orbit_model}) was plunged into the GUMS dataset. The left panels represent this mock dataset in (a) position and (c) proper motion space. The right panels represent the same dataset with the stream stars highlighted in blue. The stream is an ultra-faint feature containing only $0.015$\% of the stars in this region of sky (the total number of stars shown is $\sim 330,000$ stars). Given the variable extinction over the field, we trimmed the data below ${\rm G_0=19}$ to ensure homogeneous depth.}
\label{fig:Gaia_data_with_orbit}
\end{figure*}

The GUMS data with the mock stream model added in are shown in Figure~\ref{fig:Gaia_data_with_orbit}. This particular orbit was chosen as its position in proper motion space lies in a region of high contamination from  Galactic field stars, so that it is effectively indistinguishable from Galactic field stars. The CMD and the dependency of the proper motion errors on the G-band magnitude is shown in Figure \ref{fig:errors_on_Gaia_data_Orbit}. Comparing the number of Galactic stars to those in the mock stream ($n_{\rm stream}/n_{\rm data} \approx 0.015$\%), one can appreciate that the mock stream is an ultra-faint feature.

\subsection{\texttt{STREAMFINDER} in action}

The mock Gaia dataset is fed into the algorithm to detect the ultra-faint stream model that we have introduced into the GUMS data. We now detail the steps that the algorithm takes.

Since we are interested here in identifying halo streams, we first reduce the number of disk contaminants by rejecting those sources whose parallax differs from $1/3000 \, {\rm arcsec}$ at more than the $2\sigma$ level (i.e. objects that are likely to be closer than $3\kpc$). This makes it natural to set $3\kpc$ as the lower distance limit for analysis. To avoid having to consider objects that venture arbitrarily far, we also impose an upper distance limit in our analysis of $200\kpc$. These cuts removed 49\% of the sample.

\subsubsection{Step 1: Assigning distances based on a stellar population model}

The algorithm uses a trial SSP model of single age and metallicity to calculate the possible solutions to the absolute $M_{\rm G}$ magnitude value given the ``observed'' ${\rm G_{BP}-G_{RP}}$ colour. With old metal-poor isochrones, there are at most three absolute magnitude values ($M_{\rm G}$) possible for a given colour value. The algorithm then estimates the possible distance values ($D_i$, i = 1,2,3) of a given star based on the ``observed'' apparent G magnitude value. If at least one of the possible distance values lies within the chosen distance range $[D_{\rm min},D_{\rm max}]$, then this particular star is retained for further study. Table \ref{tab:Galaxy_range} lists the  parameter intervals that we adopted for the purpose of our analysis. By virtue of this procedure, the data that lie outside the colour range of the selected isochrone model are thrown away (in this case, leaving 42\% of the initial sample). We emphasise that this procedure does not follow from the Match Filter technique and was applied only to reduce the number of contaminants and so boost the signal to noise of the stream detection.

The algorithm then uses the derived distances of the given star along with its proper motion value to calculate the possible tangential velocities $v_t$ that it might have, corrected for solar reflex motion. Since we are interested in finding structures that are bound to the Galaxy, the total 3D velocity $v$ of the member stars of the structures must be less than the escape velocity of the Milky Way ($v_{\rm esc}$), i.e., 
\begin{equation}
 \sqrt{v_t^{2} + v_r^2} = v < v_{\rm esc} \, ,
\end{equation}
where $v_r$ is the radial velocity of the star. Since Gaia will not give us access to the entire 3D velocity of halo stars, we require only that $v_t < v_{esc}$. 

Then for a given star, which has already satisfied the distance criterion, if the condition $v_t < v_{\rm esc}$, is satisfied for any distance solution $D_i$, then this star is retained in the sample. We adopt $v_{\rm esc}=600\kms$, which corresponds to the upper limit derived by \cite{EscapeVel2007}.

\begin{figure*}
\begin{center}
\includegraphics[width=\hsize]{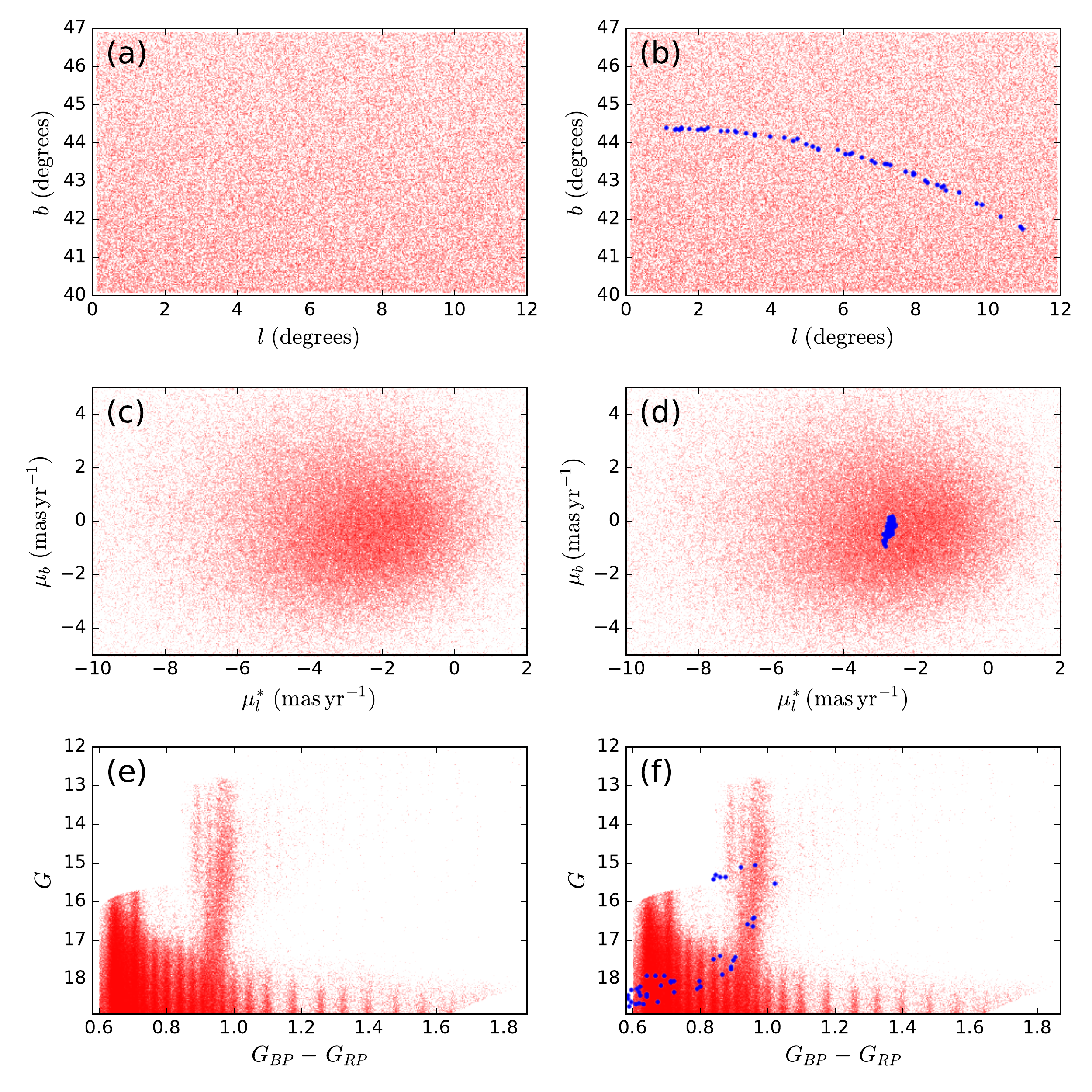}
\end{center}
\caption{Data filtering. The data are first cleaned as described in the text to remove objects with high parallaxes. Next, using the adopted SSP model to derive the distance, the algorithm rejects stars with distances and tangential velocities outside of the chosen ranges. This filtering procedure allows the algorithm to diminish the contamination from field stars, making the stream search easier. In this example, the number of stars dropped from $\approx$330,000 to $\approx$140,000. Panels (a)--(d) are identical to Figure \ref{fig:Gaia_data_with_orbit}, while panel (e) shows the Gaia CMD of the contamination and stream, with the stream highlighted in blue in panel (f).}
\label{fig:Gaia_data_Filter}
\end{figure*}

The sample after the application of these parameter cuts is shown in Figure~\ref{fig:Gaia_data_Filter}.

\begin{table}
\caption{Parameter ranges used to integrate orbits in the Galaxy.}
\label{tab:Galaxy_range}
\begin{center}
\begin{tabular}{lcc}
\hline
Parameter & $minimum$ & $maximum$  \\
\hline
$d_{\odot}$ & 3.0 kpc & 200 kpc\\
$D_{helio}$   & 3.0 kpc & 200 kpc\\
\hline
\end{tabular}
\end{center}
\end{table}

\subsubsection{Step 2: Orbit Sampling and integration}

The next task that the algorithm executes is the calculation of trial orbits for each star in the sample. Integration of trial orbits requires specifying a potential as well as the precise initial 6D phase-space position. For a given star, the algorithm has access to the 4D data astrometric information ($\ell, b, \mu_l, \mu_b$) along with the distance solutions $D_i$. The algorithm is not provided any radial velocity information (although we note that it would be trivial to include any $v_r$ measurements, if they were available). 

The proper motions have associated errors and this also does not allow us to pin-point a specific phase-space location of each star. We circumvent this issue by sampling orbits choosing parameter initial positions in the coordinates of the observables that are consistent with the corresponding uncertainty distributions. 
The on-sky 2D position measurements ($\ell, b$) are extremely accurate and hence are kept fixed. The same star has at most three possible distance values, giving three sampled distance values. Furthermore, every star has two proper motion components ($\mu_l, \mu_b$). The corresponding measurement uncertainties force us to sample values from the proper motion space as well. So for a given ($\ell, b, D_i$) combination, the algorithm samples proper motion values between $[-3\sigma_{\mu}, +3\sigma_{\mu}]$. Finally, we also sample linearly over radial velocity with a resolution of 10 $\kms$ in such a way that the total velocity covers the range $[-v_{\rm esc}, +v_{\rm esc}]$. 

\begin{figure}
\begin{center}
\includegraphics[angle=0,width=\hsize]{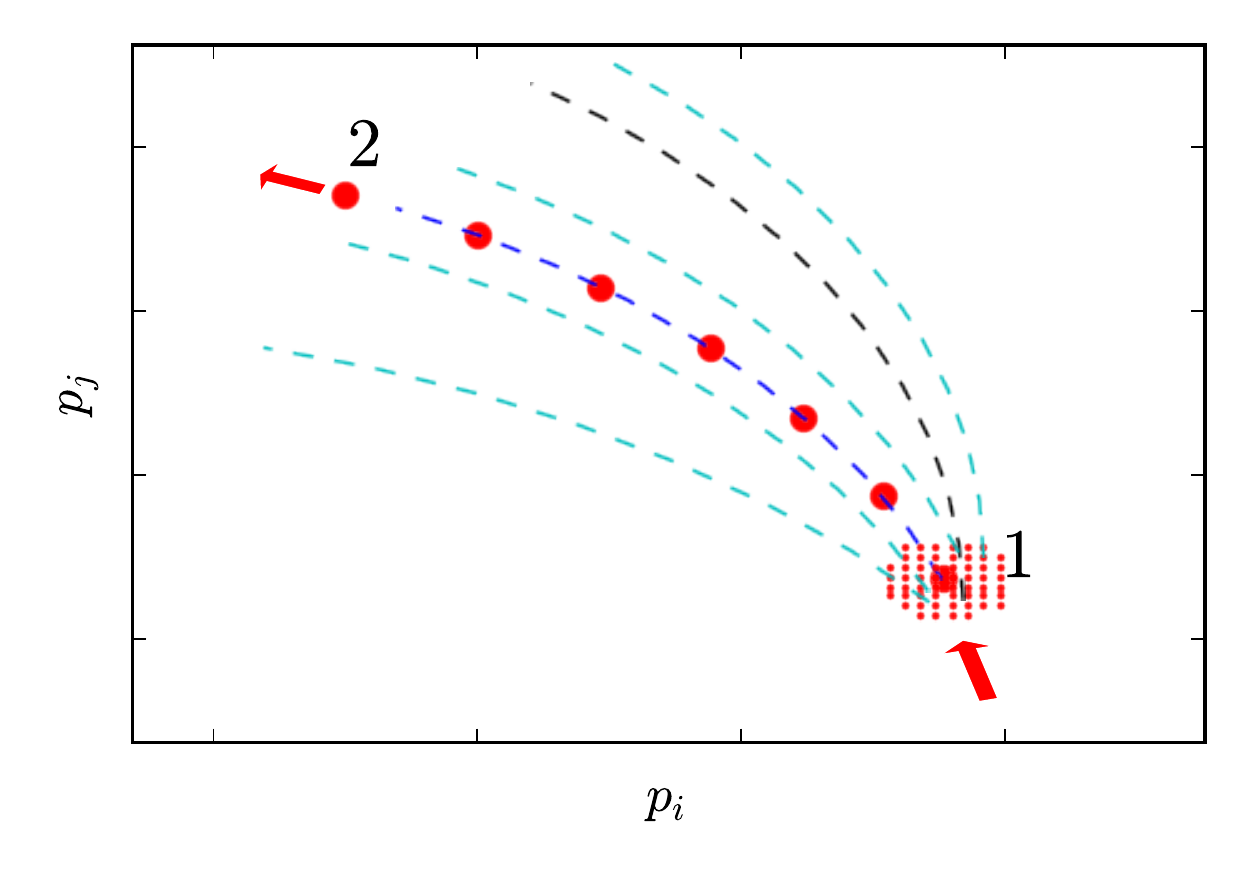}
\end{center}
\caption{Orbit sampling. Due to measurement uncertainties and the missing phase-space information of the stars, their current 6D phase-space position cannot be pinned down precisely. This uncertainty in information is illustrated here as tiny red dots around star `1' which are also the sampled phase-space positions of this star. Using these sampled phase-space positions we integrate trial orbits (cyan dashed curves) along which the streams (large red dots) are searched for in the dataset. If we had used the ``observed'' phase-space values directly for orbit-integration, instead of sampling phase-space, that might launch an orbit that is mis-aligned with the true trajectory of the star (black dashed curve) and hence may not yield a detection.}
\label{fig:orbit_sampling}
\end{figure}

In this way for every data point we get $\sim 30000$ sampled values ($n_D (\sim 3) \times n_{\mu_l} (\sim 10) \times n_{\mu_b} (\sim 10) \times n_{v_r} (\sim 100)$). Thus the uncertainty associated with the astrometric and photometric measurements, as well as the essentially completely unconstrained radial velocity, is reflected as 30,000 possible 6D positions where a given star could lie in 6D phase-space. Although this may appear to be a crude sampling of phase space, we were surprised to find that it was adequate to detect the artificial streams we simulated. To check if this given data point has other associated coherent members that share a similar orbital path, we try all of the 30,000 orbits integrated using these sampled initial conditions. The procedure is sketched in Figure~\ref{fig:orbit_sampling}.

The sampled phase-space points are integrated using a symplectic leapfrog integrator. We model the acceleration field of the Milky Way with the flexible multipole expansion software of \cite{Dehnen1998Massmodel}; for these particular tests, we again adopt their mass model 1.

\subsubsection{Algorithm Parameters}

The algorithm is provided with some generic control parameters that allow one to tune the size of the hypertube in phase-space according to the morphology of the stream structure that one aims to detect. These controls allow the algorithm to be tuned and are discussed below.

\begin{enumerate}
\item \textit{Hypertube width:} We predefine the width of the hypertubes in phase-space in terms of the allowed dispersion in the velocity space (parameter $\sigma_v$) and the allowed structural width in real space (parameter $\sigma_w$). These two parameters define the morphology of the stream that the algorithm then tries to detect. 

To make reasonable assumptions about $\sigma_w$ and $\sigma_v$, we refer to \citet{Grillmair_book_2016} and references therein, where these properties of known cold streams are listed. Based on this, we set $\sigma_w=100 \pc$ and $\sigma_v=2.0\kms$, which are appropriate for a stream derived from a low mass progenitor cluster. For comparison, this value of $\sigma_w$ when projected on the sky gives an angular width of the stream of $0.30\deg$ at $20\kpc$. For example, the GD1 stream has an angular width of $0.5\deg$ at a distance of $\sim 9\kpc$, implying a width of $70\pc$ \citep{CarlbergGD1paramter2013}. 

\item \textit{Hypertube length:} Stellar streams that are detected in Milky Way surveys have different lengths that depend on the detailed structure and mass of the progenitor, its orbit and merging history. We therefore did not fix the orbits to a particular length, but rather we integrated them until they moved out of the chosen sky window under study.
\end{enumerate}

\subsubsection{Step 3: Stream Finding}

For every trial hypertube, the algorithm tests all survey data points to establish those that are compatible with this trajectory. The orbit compatibility test is done in a 5D parameter space. Four of these dimensions come directly from the astrometry of the data in the form ($\ell, b, \mu_l, \mu_b$). The remaining dimension is one of the distance solutions $D_i$, as derived from the photometry. In practise, the algorithm uses distance moduli $DM$ to encode the distance information, in order to account easily for Gaussian uncertainties in photometry.

We model the stream as a structure that has a Gaussian distribution perpendicular to the orbit, in each of the observed dimensions of the data, and convolve this model with the corresponding observational uncertainties. For a given data point $j$, \texttt{STREAMFINDER} calculates the closest point $k$ along the trial orbit as
\begin{equation}
\omega_{\rm sky} = \sqrt{cos^2(b^j_d)(\ell^j_d - \ell^k_o)^2 + (b^j_d - b^k_o)^2} \, ,
\end{equation}
where the $\ell$ and $b$ are Galactic coordinate values and the $d$ subscript denotes ``data'', while the $o$ subscript denotes the calculated ``orbit''. If this angular distance is greater than the chosen angular model hypertube width, then this data point is considered to be incompatible with the given orbit and deemed to be a contamination star. If the datum satisfies the angular width criteria, then for the given datum $j$ and closest orbital point $k$, the algorithm calculates the following statistic, based on kinematics and structure:
\begin{equation}\label{summed_over}
\begin{split}
\mathcal{L}_{\rm kinematics} =
-\ln(\sigma_{\rm sky}\sigma_{\mu_l}\sigma_{\mu_b}\sigma_{DM})
-\frac{1}{2}\Big(  \dfrac{\omega^2_{\rm sky}}{\sigma^2_{\rm sky}}\\
+ \dfrac{(\mu^j_{l, d} - \mu^k_{l, o})^2}{\sigma^2_{\mu_l}}
+ \dfrac{(\mu^j_{b, d} - \mu^k_{b, o})^2}{\sigma^2_{\mu_b}}
+ \dfrac{(DM^j_{d} - DM^k_{o})^2}{\sigma^2_{DM}} 
\Big) \, , \\
\end{split}
\end{equation}
where $\mu^j_{l, d}, \mu^j_{b, d}$ and $DM^j_d$ are the observed proper motion and distance modulus values, and the corresponding model values are marked with the subscript $o$. As stated before, the Gaussian dispersions $\sigma_{\rm sky}, \sigma_{\mu_l}, \sigma_{\mu_b}, \sigma_{DM}$ are the convolution of the intrinsic dispersion of the model together with the observational uncertainty of each data point.

While we have constructed our statistic deliberately to resemble the logarithm of the likelihood of a model, we stress that $\mathcal{L}_{\rm kinematics}$ is not a likelihood, as that would require one to model properly the contaminating field-star population. Such modelling would be computationally very costly and hence impractical for the present purpose of {\it finding} streams.

If $\mathcal{L}_{\rm kinematics}$ is found to be greater than the floor value $\mathcal{L}_{\rm kinematics,\,floor}$ (a parameter of the algorithm), then this data point $j$ is considered to be compatible with the orbit and hence qualifies as a candidate member. The same orbit is compared to all the other stars in the dataset to find all the compatible stars. If $q$ stars out a total of $n_d$ in the survey are retained as members of the orbit, we derive the total statistic (based on the kinematics and structure), as:
\begin{equation}
L_{k} = (n_d-q)\times \mathcal{L}_{\rm kinematics,\,floor} + \sum_q \mathcal{L}_{\rm kinematics} \,.
\end{equation}
The first term on the RHS is designed to allow streams with different numbers of encapsulated stars to be compared. 

This procedure is carried out for all the trial orbits through datum $j$. The trial orbit with the highest value of $L_{\rm k}$ is considered to be the best orbit, and is then assigned to datum $j$.

\begin{figure*}
\begin{center}
\includegraphics[width=\hsize]{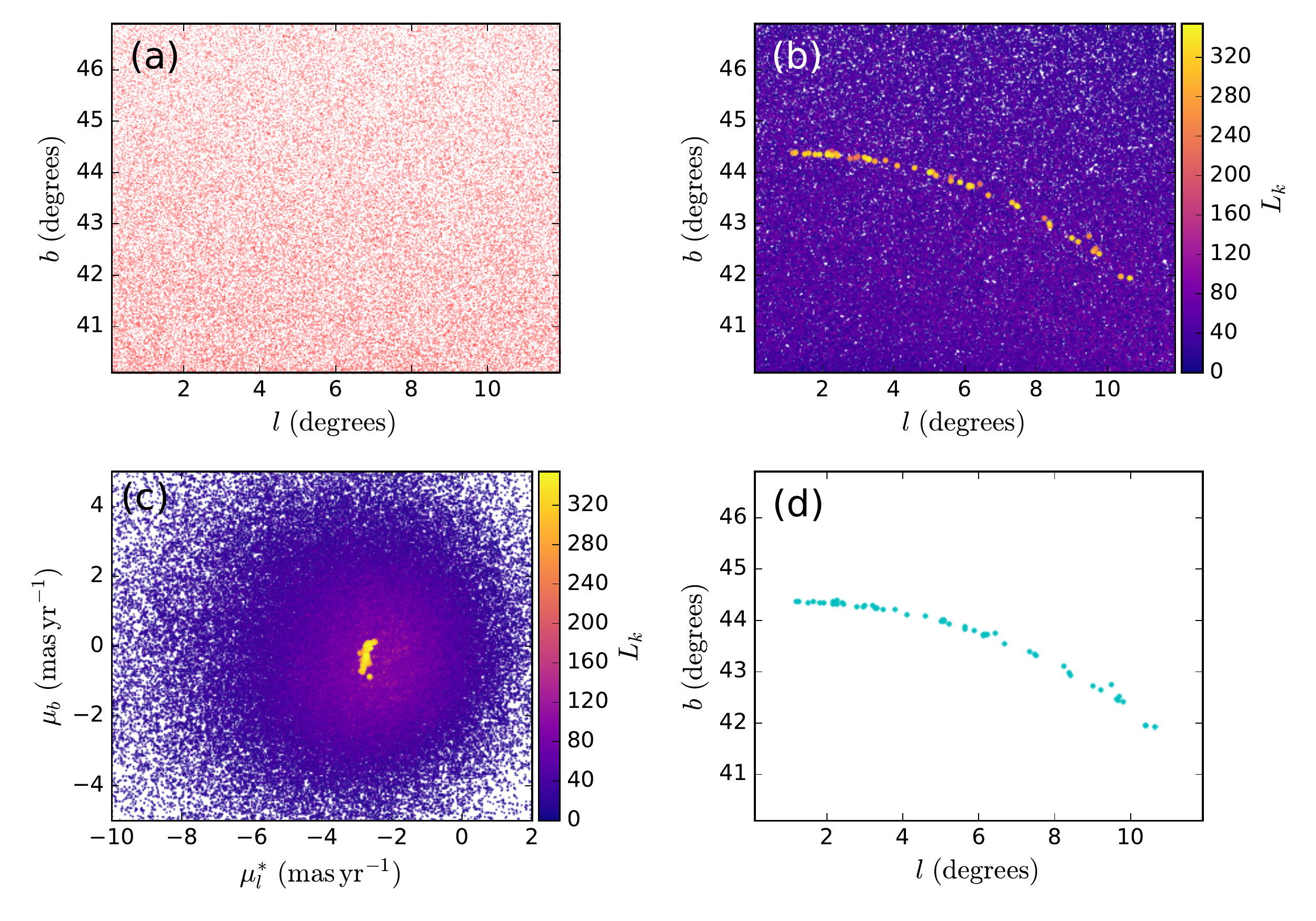}
\end{center}
\caption{\texttt{STREAMFINDER} density plot, showing the detection of an ultra-faint mock stream feature introduced into the GUMS data.  (a) The patch of sky shows no hint of the structure in density, however, it is clearly detected via the $L_k$ statistic calculated by the \texttt{STREAMFINDER} algorithm (b). The colour axis marks the relative value of $L_k$. The corresponding proper motion distribution is shown in (c). Selecting only those stars with $L_k>L_{k,{\rm max}}-150$ reveals the stream very clearly.}
\label{fig:Orbit_model_streamfinder_plot}
\end{figure*}

After processing all the data stars in this manner, the output of the algorithm can be summarized in a density plot such as that shown in Figure \ref{fig:Orbit_model_streamfinder_plot}, where the input stream model can be clearly seen. This means that despite the fact that the stream model was an ultra-faint feature, the multidimensional analysis done by \texttt{STREAMFINDER} allows it to detect even extremely low contrast objects. This procedure using orbital models as streams gives us a proof of concept of our algorithm.

\begin{figure*}
\begin{center}
\includegraphics[width=\hsize]{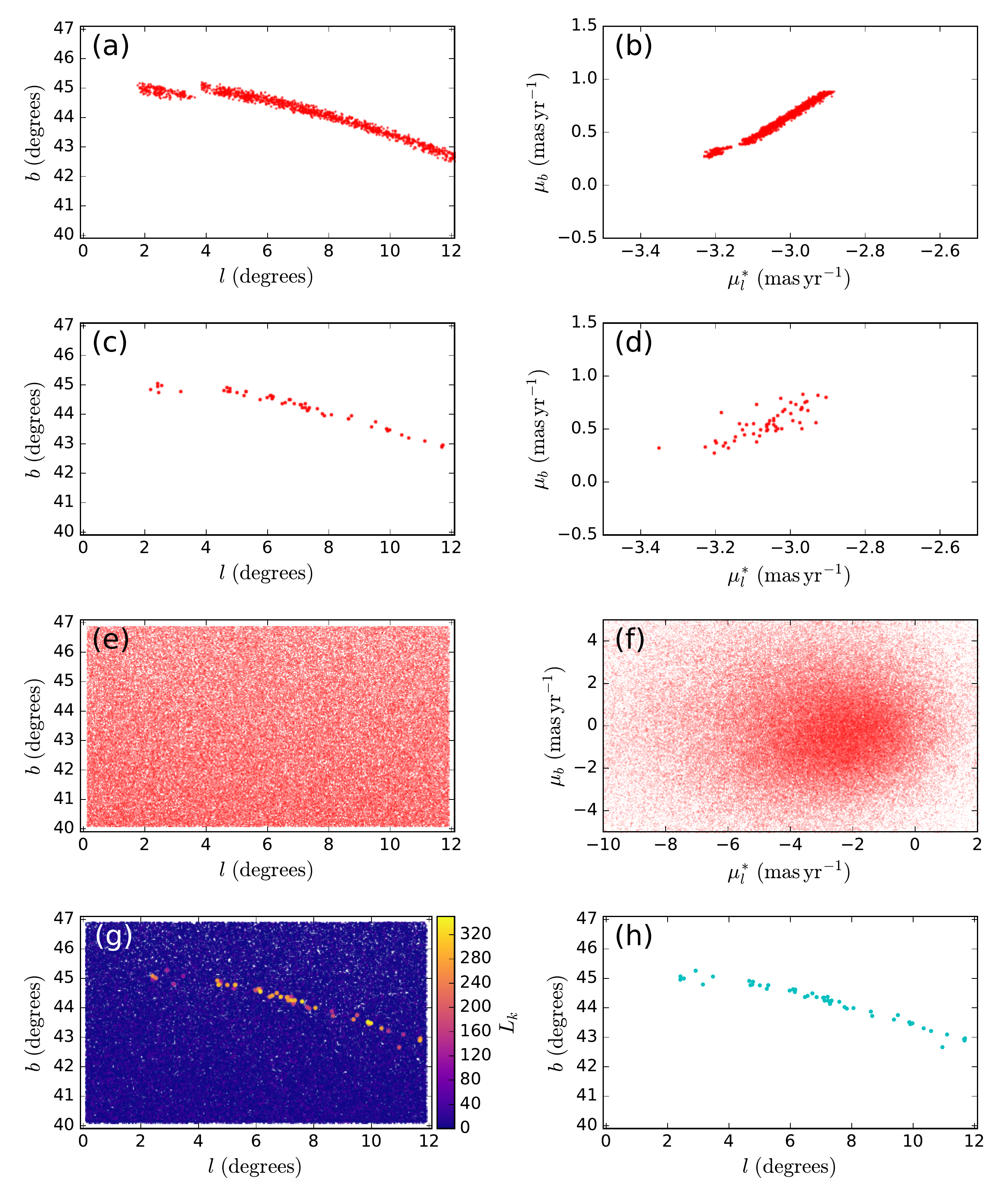}
\end{center}
\caption{N-body stream case. The top panels (a) and (b) show the simulated stream in Galactic coordinates and proper motion space. We have purposely removed the progenitor to challenge the algorithm. Figures (c) and (d) show the degraded version of the stream where the quality of the data is degraded in accordance to expected Gaia errors and only 50 data points are retained (equivalent surface brightness of $\Sigma_{\rm G} \sim 32.5\, {\rm mag \, arcsec^{-2}}$). (e) and (f) represent the GUMS data with the mock stream superposed. There are around $\sim$330000 contaminating field stars, so $n_{\rm stream}/n_{\rm data} \approx 0.015$\%. Panel (g) displays the relative likelihood $L_k$ obtained from the \texttt{STREAMFINDER}, revealing the low contrast stream feature, while (h) represents the subsample with the highest values of $L_k$.}
\label{fig:stream_sky_view}
\end{figure*}

\section{N-body simulated stream Model}\label{sec:N_body_simulated_stream_model}

In reality, star streams do not follow perfect orbits. So we next test whether our hypertube search algorithm works well with more plausible structures derived from the tidal disruption of low mass clusters. To this end we decided to produce N-body models of streams for which we used the GyrafalcON N-body integrator \citep{Dehnen2000GyrafalcON} from the NEMO software package \citep{Teuben1995NEMO}.

Although we have tested our algorithm on various mock N-body streams, we decided to present here a structure on an orbit similar to that of the Palomar 5 globular cluster stream \citep{Odenkirchen2001,Rockosi2002}. This feature is a ``poster child'' case \citep{Kupper2015} of a thin cold stream of the type that \texttt{STREAMFINDER} aims at detecting.

The mock stream was created by choosing an initial phase-space point for the progenitor cluster such that the resulting stream matches the current position, distance, and extension of the Pal 5 stream. The progenitor was built using a King model (\citealt{King1966}), with mass, tidal radius and ratio between central potential and velocity dispersion of $M_{\rm sat} = 2 \times 10^{4}\msun$ , $r_{t} = 50\pc$ and $W_{\rm sat} =2.5$, respectively \citep{Thomas2016}. Once the progenitor was initialised in phase space, it was then evolved forwards for $3.0\Gyr$ in the adopted Galactic mass model. In order to make the detection more challenging, at the end of the simulation we removed the stars within $50\pc$ from the progenitor remnant from the sample. Our N-body stream closely follows the structure and kinematics of the true Pal~5 stream, though we stress that the purpose here is not to make a quantitative comparison with the real stellar structure. 

A similar procedure as before was followed to assign Gaia-like proper motion uncertainties and Gaia colour-magnitude values to the N-body particles. The degraded version of the simulated stream was immersed in the same degraded contamination (GUMS) model as used previously in Section~\ref{sec:Orbital_Stream_models}. The simulated data with the mock N-body stream immersed in it is shown in Figure \ref{fig:stream_sky_view}. We chose to incorporate only 50 stream stars in this test ($<4$\% of the probable $2 \times 10^{4}\msun$ progenitor of Pal~5), which amounts to $0.015$\% of the sample. The equivalent surface brightness of the mock stream candidate is $\Sigma_{\rm G} \sim 32.5\, {\rm mag \, arcsec^{-2}}$. 

This data was then fed to the \texttt{STREAMFINDER} algorithm to detect this \textit{ultra-faint} stream feature following exactly the same procedure and analysis as described in Section~\ref{sec:Orbital_Stream_models}. The output of the algorithm is the map of the stream $L_k$ statistic shown on the bottom panels of Figure \ref{fig:stream_sky_view}. The stream members can be clearly identified above the contamination in this map. 

This case-study demonstrates the success of our algorithm in detecting realistic and extremely faint stream features in a Gaia-like dataset.

\begin{figure*}
\begin{center}
\includegraphics[width=\hsize]{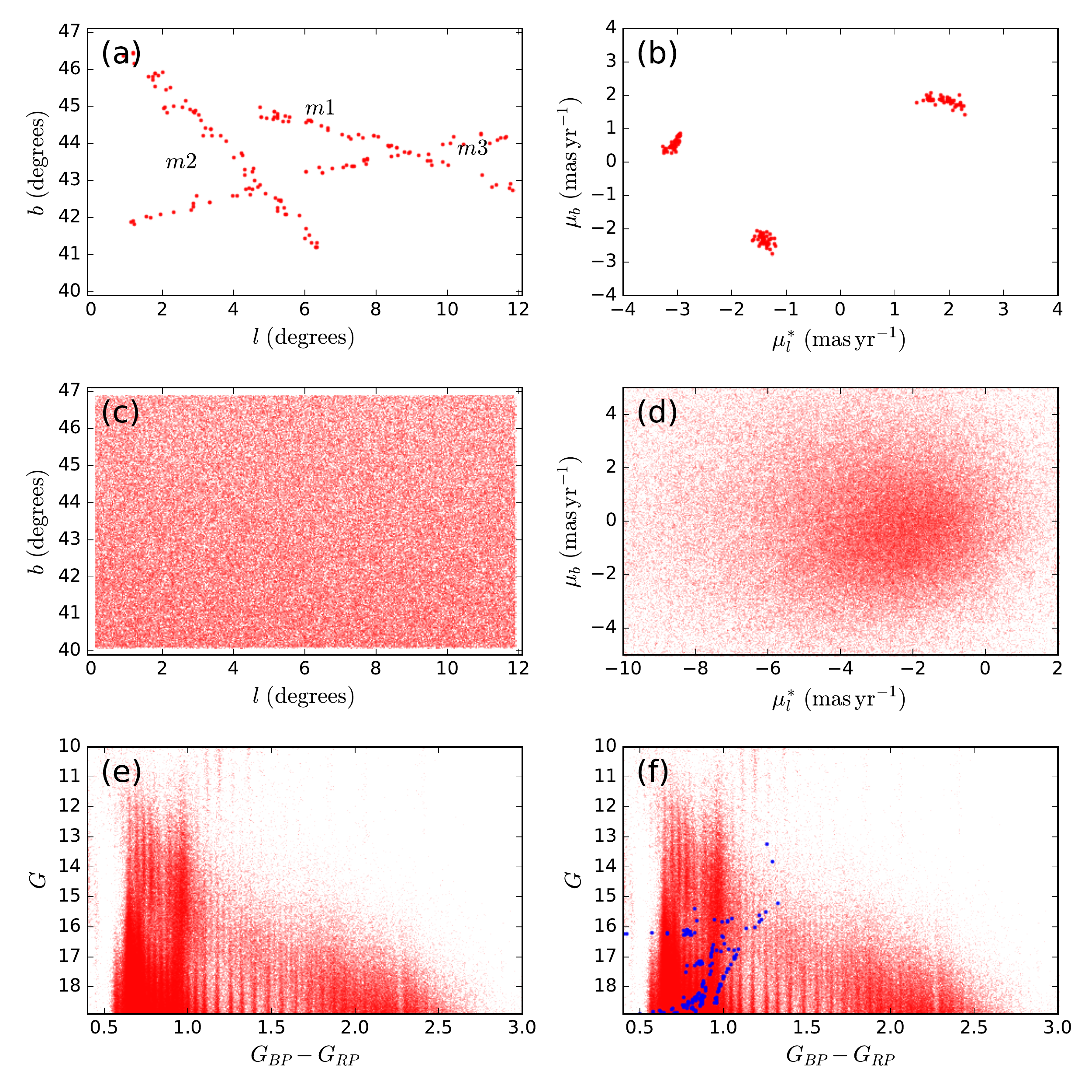}
\end{center}
\caption{Multiple stream case. The top panels show the degraded version of the three N-body simulated streams in Galactic coordinates. The middle panels show the GUMS data with the three streams immersed. The bottom panels show the colour-magnitude distribution of these data: (e) shows the data along with the streams within it, while in (f) the stream is highlighted in blue. We chose 3 isochrone models appropriate for halo globular clusters with age and ${\rm [Fe/H]} = (10\Gyr,-1.28), (10\Gyr, -1.58)$ and $(10\Gyr, -2.28)$ for, respectively, models m1, m2 and m3. Though not explicitly shown here, the streams probe distances between 10 to $28\kpc$. Each stream possesses 50 stars, and has an equivalent surface brightness of $\Sigma_{\rm G} \sim 32.5\, {\rm mag \, arcsec^{-2}}$.}
\label{fig:multiple_stream_sky_view}
\end{figure*}

\begin{figure*}
\begin{center}
\includegraphics[width=\hsize]{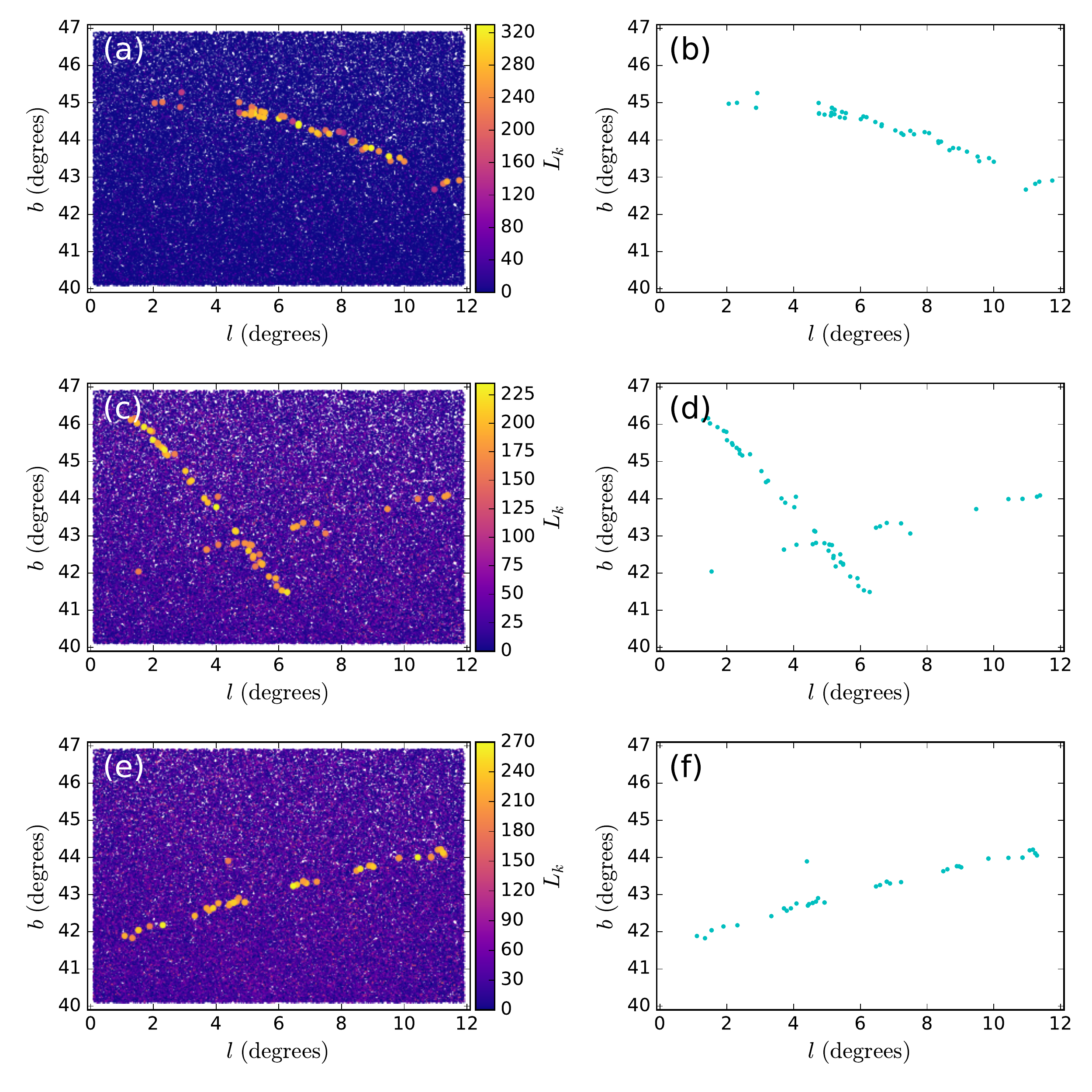}
\end{center}
\caption{\texttt{STREAMFINDER} results for the case of multiple streams in  a given patch of sky. The left panels show the spatial distribution of the statistic $L_k$ obtained using different isochrone models, and the right panels show the data points with the highest $L_k$ values. The upper, middle and lower panels are derived using, respectively, the SSP models with age and ${\rm [Fe/H]} = (10\Gyr,-1.28), (10\Gyr, -1.58)$ and $(10\Gyr, -2.28)$. As expected, a given isochrone model enhances the detection strength of the stream structures corresponding to that particular isochrone. (The $L_k$ values shown here are values relative to the minimum).}
\label{fig:Multiple_stream_density_plot}
\end{figure*}

\section{Multiple Streams}
\label{sec:Multiple_Streams}

In the hierarchical picture of galaxy formation, galaxies like the Milky Way grow by repeated merging and accretion of their satellites. Some of the disrupted satellites will have contained star clusters \citep{Ibata2003Bellazzini}, which themselves will eventually tidally disrupt to form long streams in the Galactic halo. The ``Field-of-Streams'' image presented by \cite{Belokurov2006} and the halo substructures map created by \cite{Bernard2016}, both show a Galactic sky full of stream-like sub-structures. These images, along with the many other detections of streams over the past few years, strongly suggests that a significant fraction of the stellar halo population is a result of hierarchical merging. As the timescales for phase-mixing are extremely long, it may turn out the Milky Way halo is a patchwork of criss-crossing streams. This may be verified once Gaia DR2 delivers its excellent astrometric solutions for the stars over the entire sky.

Therefore, we also test the ability of our algorithm to make detections in this much more interesting case where a patch of sky contains multiple streams laid over each other. For this test, we again use the \citet{Dehnen1998Massmodel} mass model 1 and the GyrafalcOn N-body integrator to produce mock streams. We chose to model three such structures. We keep the same (Palomar~5-like) mock stream as previously, and add two new random streams.

The initial phase space distribution of the three progenitors of the streams were selected as follows. The initial position of each satellite was drawn at a random direction as seen from the Galactic centre, and with a uniform probability of lying in the Galactocentric distance range of $[10\textup{--}30]\kpc$. The mean velocity of each satellite was selected randomly from an isotropic Gaussian distribution with (one-dimensional) dispersion of $100\kms$ \citep{Harris1976}. At these phase space positions, each progenitor was constructed using a King model (\citealt{King1966}). The mass, tidal radius and ratio between central potential and velocity dispersion were sampled uniformly between the ranges $M_{sat} = [2\textup{--}4] \times 10^{4}\msun$ , $r_{t} = [20\textup{--}80]\pc$ and $W_{sat} = [2\textup{--}4]$. 

Once the progenitors were initialised in phase space, they were then evolved independently over a time period between $[2\textup{--}6]\Gyr$ in the same Galactic mass model mentioned above. We re-sampled the initial conditions of those progenitors that did not disrupt or did not fall into the chosen sky region. Each of the three streams was assigned an SSP isochrone model of age and metallicity $(10\Gyr,-1.28)$, $(10\Gyr, -1.58)$ and $(10\Gyr, -2.28)$, which cover plausible values for halo globular clusters. These streams were degraded in their astrometric measurements and were introduced into a common contamination model in the same manner as in Section~\ref{sec:Orbital_Stream_models}. The data provided to the algorithm is shown in Figure \ref{fig:multiple_stream_sky_view}. 

The algorithm was re-run three times with these data, each time using one of the three isochrone models to assign distances to the stream stars. Figure \ref{fig:Multiple_stream_density_plot} shows the resulting stream maps, where the first row uses the correct stellar populations model for stream m1, the middle row for stream m2, and the third row for stream m3.

This shows that the procedure needs to use the correct trial SSP model to successfully detect the input streams. With the real Gaia data it will be necessary to run the algorithm over a grid in metallicity and age (our tests suggest that intervals of $0.1$~dex and $\sim 1\Gyr$ are appropriate).

\section{Luminosity function and continuity: additional \texttt{STREAMFINDER} criteria} 
\label{STREAMFINDER_tests}

So far, we have discussed searching for sub-groups of stars in a sample whose kinematic and spatial properties mirror a plausible orbit. We will now also include two additional criteria that will help improve further the contrast of faint structures. 

Our algorithm aims to find thin and cold stream structures. These structures are expected to be remnants of a globular cluster and are formed by their disruption and dissolution. The member stars of most star clusters follow closely stellar evolutionary models of a single age and metallicity, and although now totally disrupted, the stream stars share similar age and metallicity as that of the progenitor and hence must follow a similar isochrone track. We incorporate this concept into our algorithm, thus making use of the photometric information of each candidate group of stars identified by the algorithm.

To this end we use as a template the G-band cumulative luminosity function of the same SSP model as was used to derive the distance solution with the hypertube technique. For each candidate group of stars, we calculate the Kolmogorov-Smirnoff Test probability $P_{\rm KS, LF}$ that the stars are drawn from this model luminosity function.

We further expect that stellar streams are extended structures, yet so far the criteria that have been described do not allow us to distinguish an extended stream from a small-scale localized over-density. To remedy this, we incorporated an additional criterion into the algorithm to calculate the Kolmogorov-Smirnoff Test probability $P_{\rm KS, continuous}$ that the member stars of a candidate structure are uniformly distributed along the orbit segment contained within the sky window under study. 

The final statistic we use is then:
\begin{equation}
L = L_{k} + \ln(P_{\rm KS, LF}) + \ln(P_{\rm KS, continuous}) \, .
\end{equation}

\begin{figure*}
\begin{center}
\includegraphics[width=\hsize]{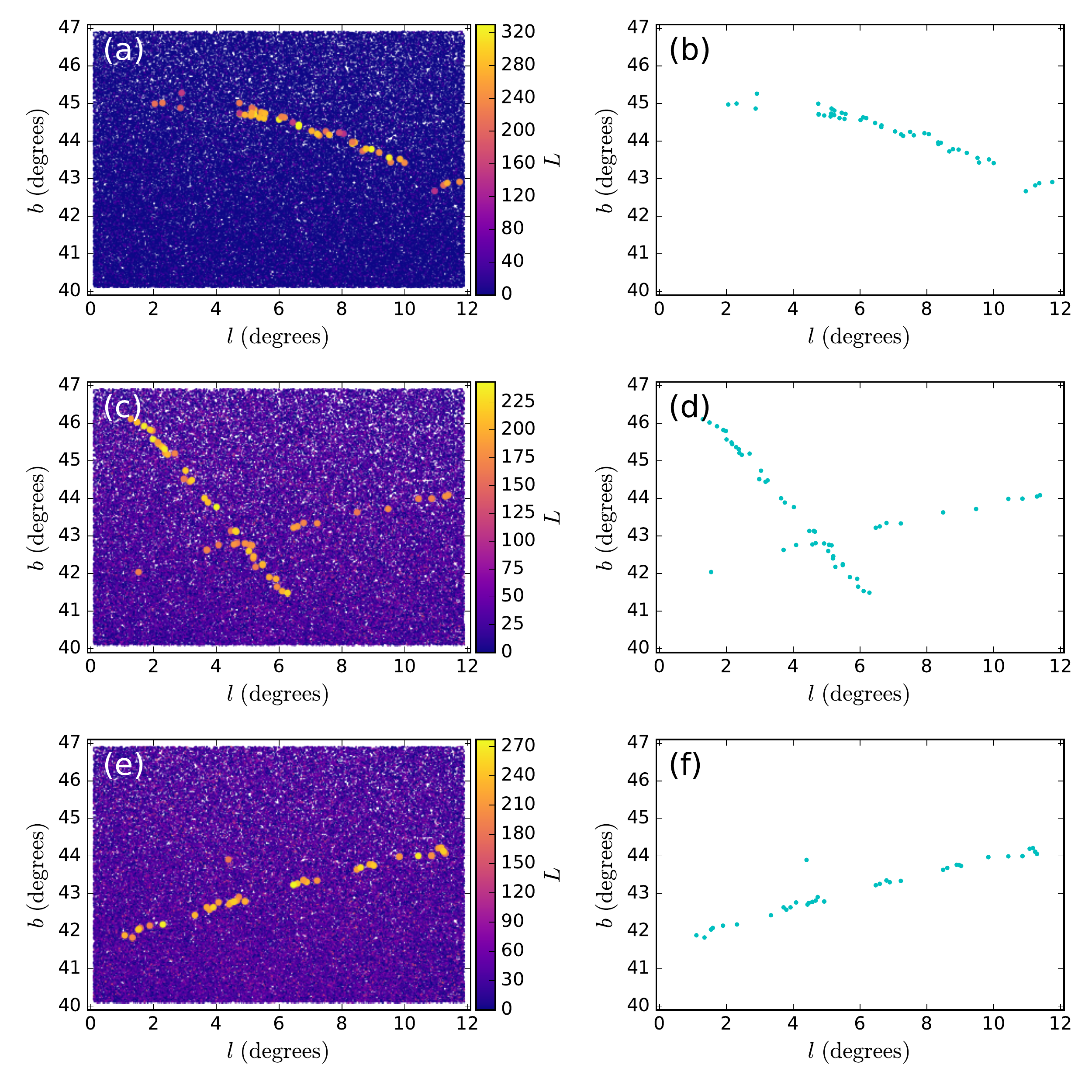}
\end{center}
\caption{Luminosity function and continuity criteria. These plots are improved versions of those shown in Figure \ref{fig:Multiple_stream_density_plot}, after incorporating the luminosity function and the continuity criteria in the likelihood calculation. The contrast of the streams is further improved by the additional discriminating information.}
\label{fig:Multiple_stream_final_density_plot}
\end{figure*}

Figure \ref{fig:Multiple_stream_final_density_plot} is an improved version of Figure \ref{fig:Multiple_stream_density_plot} after incorporating the luminosity function and the continuity criteria into the $L$ statistic used by the algorithm. As can be seen by comparing the colour axes of the two figures, the additional criteria improve the contrast of the detection.

\section{Testing the Detection Limit}\label{sec:Det_limit}

\begin{figure*}
\begin{center}
\includegraphics[width=\hsize]{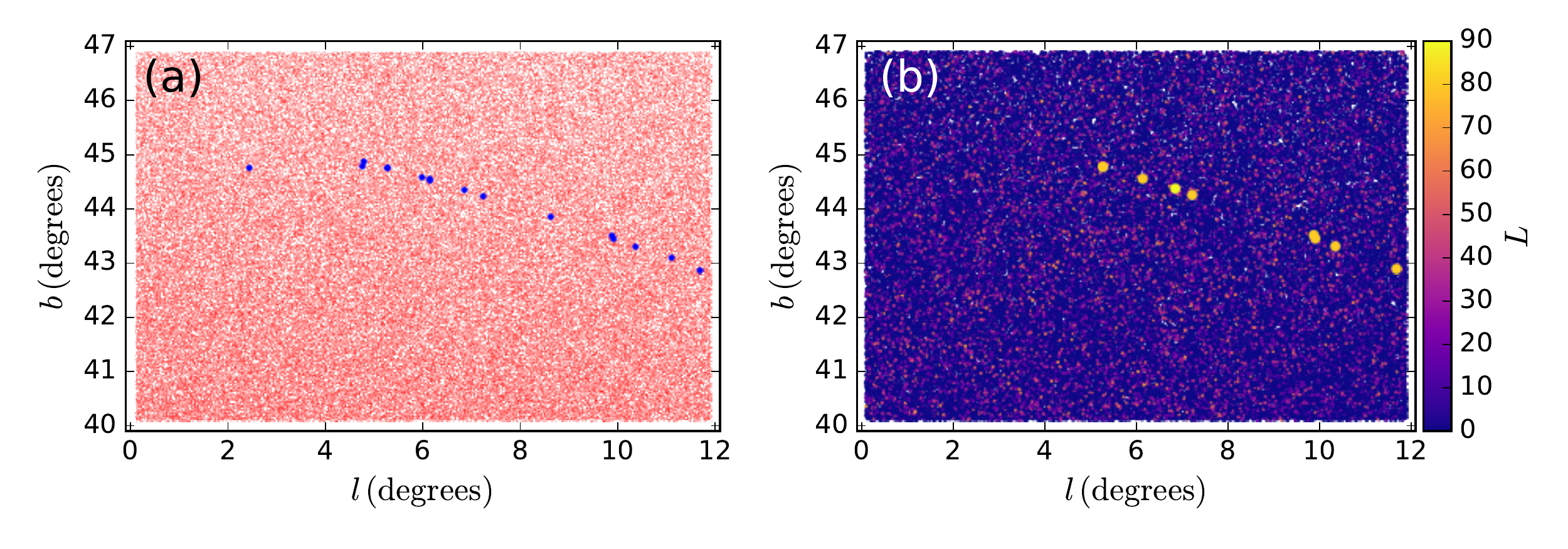}
\end{center}
\caption{Detection limit test. Here, we have simulated an ultra-faint stream structure possessing only 15 stars (with $\Sigma_{\rm G} \sim 33.6\, {\rm mag \, arcsec^{-2}}$). The spatial distribution is shown in (a) where the input stream stars are highlighted in blue, while (b) shows the corresponding $L$ statistic: 10 stars are clearly detected above a $\sim 4.3\sigma$ threshold.}
\label{fig:detection_limit_plot}
\end{figure*}

It is useful to gauge the faintest stream structure (in terms of number of stream stars) the algorithm can detect. To this end we reran our algorithm over the m1 mock dataset, which shares the orbital properties of the Palomar~5 globular cluster. We reran the algorithm, removing one star at a time from the stream to see at what point the structure becomes lost in the noise. We found that with an initial stream containing 15 stars, 10 were recovered with values of the $L$ statistic higher than 1 in 150000 among the contaminating population (i.e., $\sim 4.3\sigma$). The corresponding stream has an equivalent surface brightness of $\Sigma_{\rm G} \sim 33.6\, {\rm mag \, arcsec^{-2}}$ over this $>10\deg$ region, and is shown in Figure \ref{fig:detection_limit_plot}. This is very promising and means that the application of our algorithm onto the actual Gaia dataset could reveal the presence of ultra-faint streams.

We must point out that this limit depends on the number of contaminants, the observational errors and on the morphology of the structures that are present in the halo. However, the test case that we simulated here shares the orbit of the real Palomar~5 (albeit with a much lower surface brightness), and so we think it provides a useful preview of the detectability of a very tenuous stream at an advanced stage of tidal disruption.

\section{Effect of adopting a wrong Galactic potential}\label{sec:DB_model4}

Hitherto we have presented test cases where the trial orbits were integrated in the same Galactic potential model in which the mock streams were originally simulated. Although the \cite{Dehnen1998Massmodel} mass model 1 we have employed here was a reasonable fit to available data in 1998, the Milky Way potential may in reality be fairly different.

\begin{figure*}
\begin{center}
\includegraphics[width=\hsize]{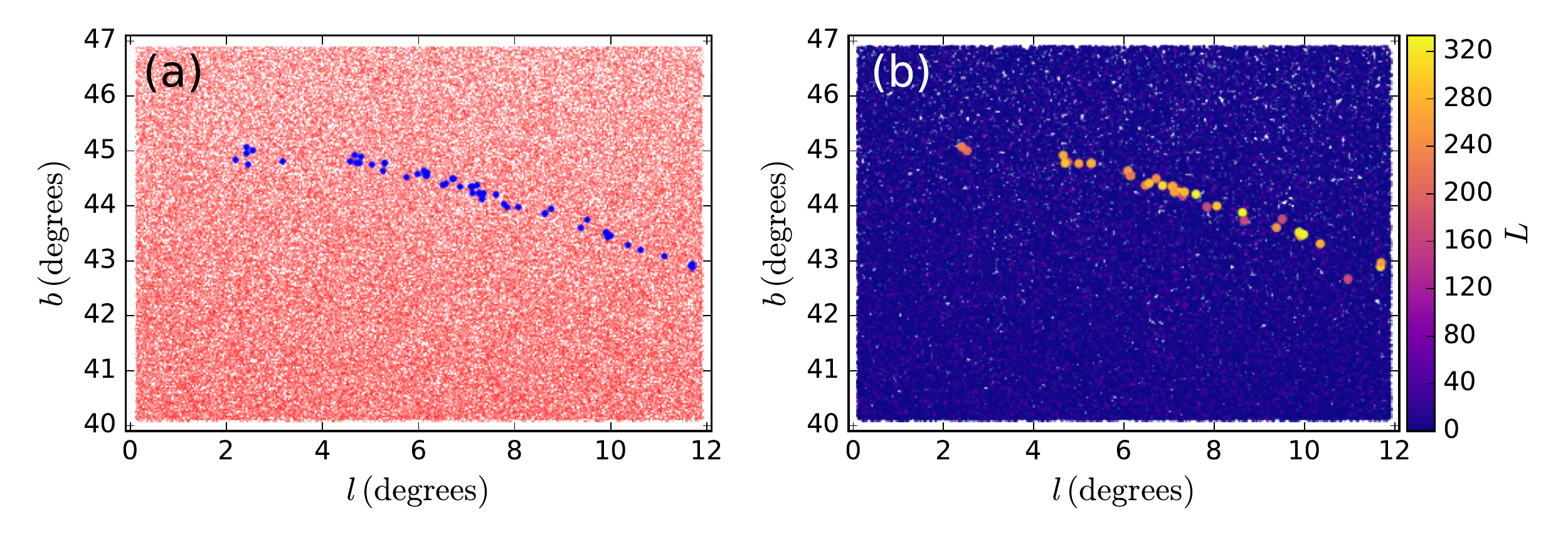}
\end{center}
\caption{Galaxy mass model mismatch. (a) shows the same data and the superposed stream model as shown in Figure \ref{fig:stream_sky_view}. The stream model was simulated in DB model 1. (b) shows the corresponding $L$ statistic obtained by using DB model 4 for integration of the trial orbits in the \texttt{STREAMFINDER} algorithm. It can be seen that the algorithm was easily able to detect the stream even after we forced the code to employ a wrong Galactic potential model.}
\label{fig:DB_model_4_plot}
\end{figure*}

To gauge the effect of adopting a wrong mass model, we reran the \texttt{STREAMFINDER} on exactly the same stream as shown previously in Section \ref{sec:N_body_simulated_stream_model}, but this time we incorporated the \cite{Dehnen1998Massmodel} mass model 4 in the detection algorithm.  The resulting distribution of the statistic $L$ is shown in Figure \ref{fig:DB_model_4_plot}, which can be seen to be similar to the counterpart in Figure \ref{fig:stream_sky_view}. 

We suspect that by iterating over different mass models it should be possible to find the potential that maximises the contrast of stellar streams in the Milky Way. However, we would like to stress that the \texttt{STREAMFINDER} is intended as an initial detection tool. Once a sample of streams have been found, we intend to use other more accurate methods (e.g. N-body simulations) to model them.

\section{Discussion and Conclusions}\label{sec:Conclusion}

In this contribution we have presented the \texttt{STREAMFINDER}, a new algorithm that aims to efficiently detect stellar stream-like features. It has been optimised to identify very faint structures using data of the quality that will soon be delivered by Gaia DR2. At its heart, \texttt{STREAMFINDER} shoots trial orbits within a realistic Galactic potential, using the astrometric and photometric measurements of the stars to select initial conditions for the orbits. These orbits are then adjusted to find the local maxima in star counts that are compatible with the trial orbit in 2D position and 2D kinematics (nevertheless, the algorithm can be easily modified to explore the full 6D phase-space information available for any sub-sample of the data). 

Every star is assigned a likelihood value based on how coherent it is with an extended stellar stream. Our tests using N-body simulated streams superimposed on the GUMS dataset with kinematics degraded to Gaia DR2-like quality and precision show that the algorithm can detect structures lying well below previous detection limits. Because our method relies on detecting stream candidates along orbits, the algorithm can detect structures that lie along radial or other complex trajectories. 

The algorithm returns a statistic that is similar to a likelihood, which must be calibrated locally to determine the structure significance, as it depends on the (varying) ``background'' population. The expected distribution of the statistic in the absence of a stream-structure may be estimated via the application of the \texttt{STREAMFINDER} to artificial data (such as the GUMS simulation) or completely empirically via the examination of the behaviour of the statistic in neighbouring regions of sky.

The design of the algorithm is such that along with the stream detection, it renders other useful insights about the detected candidate structures that can be used for further analysis.
\begin{enumerate}
\item The algorithm delivers the orbital structure along which the stream lies: This is the primary by-product that the algorithm naturally returns and gives the possible set of orbital solutions that the stream might lie along. Radial velocities and distance information of the stars will be missing for the great majority of halo stars in the Gaia DR2 (and later) catalogues. However, since the algorithm gives the possible orbital solutions for a given stream structure, it therefore provides a means to complete the 6D phase-space solutions that are possible for a given stream star.
\item Phase-space distribution of streams: The algorithm delivers a complete 6D phase-space distribution of possible orbital solutions that a given stream might be on. When executed over the entire sky, the end product would be the distribution function of stream stars in the Galactic halo. This solution could be extremely useful for re-creating the pre-merging history of the Milky Way, or to perform Schwarzschild modelling to constrain the dark matter distribution in the Galaxy.
\item The SSP test is intrinsically incorporated into the algorithm: Most of the coherence-based detection schemes do not always take into account the best suited stellar population model for the candidate stream structure. However, our algorithm calculates the likelihood of every stream candidate based on  SSP models, thus our approach also returns a possible set of SSP models that the stream might correspond to. This can be viewed as a low-resolution ``chemical tagging'' approach, where stars can be tagged based on their age and metallicities giving an orbit-age-metallicity distribution of stars in the Milky Way halo.
\item Length of the structure: The algorithm also allows us to estimate the linear length of the candidate structures simply by summing along the orbit until some lower detection threshold is reached. Through subsequent modelling, this can be converted into an estimate of the minimum age since the disruption of the progenitor.
\item Calculating orbital properties: Since the algorithm offers orbital solutions for every stream, one can easily calculate simple orbital properties of the stream structure such as the eccentricity or energy of the streams.
\end{enumerate}

Motivated by these results, and to test the machinery on real data, we have applied it to the Pan-STARRS1 dataset \citep{PanSTARRS_Kaiser2002, Chambers2016PS1, PanSTARRS2016}, the results of which will be presented in the next contribution in this series (Malhan et al. 2018, in prep.).

\section*{Acknowledgements}
The authors would like to thank the anonymous referee for very useful comments that contributed to the clarity and overall improvement of the paper.


\bibliographystyle{mnras}
\bibliography{ref1} 

\begin{thebibliography}{}
\makeatletter
\relax
\def\mn@urlcharsother{\let\do\@makeother \do\$\do\&\do\#\do\^\do\_\do\%\do\~}
\def\mn@doi{\begingroup\mn@urlcharsother \@ifnextchar [ {\mn@doi@}
  {\mn@doi@[]}}
\def\mn@doi@[#1]#2{\def\@tempa{#1}\ifx\@tempa\@empty \href
  {http://dx.doi.org/#2} {doi:#2}\else \href {http://dx.doi.org/#2} {#1}\fi
  \endgroup}
\def\mn@eprint#1#2{\mn@eprint@#1:#2::\@nil}
\def\mn@eprint@arXiv#1{\href {http://arxiv.org/abs/#1} {{\tt arXiv:#1}}}
\def\mn@eprint@dblp#1{\href {http://dblp.uni-trier.de/rec/bibtex/#1.xml}
  {dblp:#1}}
\def\mn@eprint@#1:#2:#3:#4\@nil{\def\@tempa {#1}\def\@tempb {#2}\def\@tempc
  {#3}\ifx \@tempc \@empty \let \@tempc \@tempb \let \@tempb \@tempa \fi \ifx
  \@tempb \@empty \def\@tempb {arXiv}\fi \@ifundefined
  {mn@eprint@\@tempb}{\@tempb:\@tempc}{\expandafter \expandafter \csname
  mn@eprint@\@tempb\endcsname \expandafter{\@tempc}}}

\bibitem[\protect\citeauthoryear{{Arifyanto} \& {Fuchs}}{{Arifyanto} \&
  {Fuchs}}{2006}]{Arcturus_Arifyanto_2006}
{Arifyanto} M.~I.,  {Fuchs} B.,  2006, \mn@doi [\aap]
  {10.1051/0004-6361:20054355}, \href
  {http://adsabs.harvard.edu/abs/2006A%26A...449..533A} {449, 533}

\bibitem[\protect\citeauthoryear{{Balbinot}, {Santiago}, {da Costa}, {Makler}
  \& {Maia}}{{Balbinot} et~al.}{2011}]{Balbinot2011MF}
{Balbinot} E.,  {Santiago} B.~X.,  {da Costa} L.~N.,  {Makler} M.,   {Maia}
  M.~A.~G.,  2011, \mn@doi [\mnras] {10.1111/j.1365-2966.2011.19044.x}, \href
  {http://adsabs.harvard.edu/abs/2011MNRAS.416..393B} {416, 393}

\bibitem[\protect\citeauthoryear{{Bellazzini}, {Ferraro}  \&
  {Ibata}}{{Bellazzini} et~al.}{2003}]{Ibata2003Bellazzini}
{Bellazzini} M.,  {Ferraro} F.~R.,   {Ibata} R.,  2003, \mn@doi [\aj]
  {10.1086/344072}, \href {http://adsabs.harvard.edu/abs/2003AJ....125..188B}
  {125, 188}

\bibitem[\protect\citeauthoryear{{Belokurov} et~al.,}{{Belokurov}
  et~al.}{2006}]{Belokurov2006}
{Belokurov} V.,  et~al., 2006, \mn@doi [\apjl] {10.1086/504797}, \href
  {http://adsabs.harvard.edu/abs/2006ApJ...642L.137B} {642, L137}

\bibitem[\protect\citeauthoryear{{Bernard} et~al.,}{{Bernard}
  et~al.}{2016}]{Bernard2016}
{Bernard} E.~J.,  et~al., 2016, \mn@doi [\mnras] {10.1093/mnras/stw2134}, \href
  {http://adsabs.harvard.edu/abs/2016MNRAS.463.1759B} {463, 1759}

\bibitem[\protect\citeauthoryear{{Bovy}, {Bahmanyar}, {Fritz}  \&
  {Kallivayalil}}{{Bovy} et~al.}{2016}]{Bovy2016GD1Pal5}
{Bovy} J.,  {Bahmanyar} A.,  {Fritz} T.~K.,   {Kallivayalil} N.,  2016, \mn@doi
  [\apj] {10.3847/1538-4357/833/1/31}, \href
  {http://adsabs.harvard.edu/abs/2016ApJ...833...31B} {833, 31}

\bibitem[\protect\citeauthoryear{{Carlberg} \& {Grillmair}}{{Carlberg} \&
  {Grillmair}}{2013}]{CarlbergGD1paramter2013}
{Carlberg} R.~G.,  {Grillmair} C.~J.,  2013, \mn@doi [\apj]
  {10.1088/0004-637X/768/2/171}, \href
  {http://adsabs.harvard.edu/abs/2013ApJ...768..171C} {768, 171}

\bibitem[\protect\citeauthoryear{{Carlberg}, {Grillmair}  \&
  {Hetherington}}{{Carlberg} et~al.}{2012}]{StreamGap_Carlberg2012}
{Carlberg} R.~G.,  {Grillmair} C.~J.,   {Hetherington} N.,  2012, \mn@doi
  [\apj] {10.1088/0004-637X/760/1/75}, \href
  {http://adsabs.harvard.edu/abs/2012ApJ...760...75C} {760, 75}

\bibitem[\protect\citeauthoryear{{Chambers} et~al.,}{{Chambers}
  et~al.}{2016a}]{Chambers2016PS1}
{Chambers} K.~C.,  et~al., 2016a, preprint, \href
  {http://adsabs.harvard.edu/abs/2016arXiv161205560C} {} (\mn@eprint {arXiv}
  {1612.05560})

\bibitem[\protect\citeauthoryear{{Chambers} et~al.,}{{Chambers}
  et~al.}{2016b}]{PanSTARRS2016}
{Chambers} K.~C.,  et~al., 2016b, preprint, \href
  {http://adsabs.harvard.edu/abs/2016arXiv161205560C} {} (\mn@eprint {arXiv}
  {1612.05560})

\bibitem[\protect\citeauthoryear{{Dalal} \& {Kochanek}}{{Dalal} \&
  {Kochanek}}{2002}]{Dalal2002}
{Dalal} N.,  {Kochanek} C.~S.,  2002, \mn@doi [\apj] {10.1086/340303}, \href
  {http://adsabs.harvard.edu/abs/2002ApJ...572...25D} {572, 25}

\bibitem[\protect\citeauthoryear{{Dehnen}}{{Dehnen}}{2000}]{Dehnen2000GyrafalcON}
{Dehnen} W.,  2000, \mn@doi [\apjl] {10.1086/312724}, \href
  {http://adsabs.harvard.edu/abs/2000ApJ...536L..39D} {536, L39}

\bibitem[\protect\citeauthoryear{{Dehnen} \& {Binney}}{{Dehnen} \&
  {Binney}}{1998}]{Dehnen1998Massmodel}
{Dehnen} W.,  {Binney} J.,  1998, \mn@doi [\mnras]
  {10.1046/j.1365-8711.1998.01282.x}, \href
  {http://adsabs.harvard.edu/abs/1998MNRAS.294..429D} {294, 429}

\bibitem[\protect\citeauthoryear{{Dehnen}, {Odenkirchen}, {Grebel}  \&
  {Rix}}{{Dehnen} et~al.}{2004}]{Dehnen2004thinorbit}
{Dehnen} W.,  {Odenkirchen} M.,  {Grebel} E.~K.,   {Rix} H.-W.,  2004, \mn@doi
  [\aj] {10.1086/383214}, \href
  {http://adsabs.harvard.edu/abs/2004AJ....127.2753D} {127, 2753}

\bibitem[\protect\citeauthoryear{{Duffau}, {Zinn}, {Vivas}, {Carraro},
  {M{\'e}ndez}, {Winnick}  \& {Gallart}}{{Duffau}
  et~al.}{2006}]{Virgo_Duffau_2006}
{Duffau} S.,  {Zinn} R.,  {Vivas} A.~K.,  {Carraro} G.,  {M{\'e}ndez} R.~A.,
  {Winnick} R.,   {Gallart} C.,  2006, \mn@doi [\apjl] {10.1086/500130}, \href
  {http://adsabs.harvard.edu/abs/2006ApJ...636L..97D} {636, L97}

\bibitem[\protect\citeauthoryear{{Erkal}, {Belokurov}, {Bovy}  \&
  {Sanders}}{{Erkal} et~al.}{2016}]{StreamGap_Erkal2016}
{Erkal} D.,  {Belokurov} V.,  {Bovy} J.,   {Sanders} J.~L.,  2016, \mn@doi
  [\mnras] {10.1093/mnras/stw1957}, \href
  {http://adsabs.harvard.edu/abs/2016MNRAS.463..102E} {463, 102}

\bibitem[\protect\citeauthoryear{{Eyre} \& {Binney}}{{Eyre} \&
  {Binney}}{2009}]{EyreBinney2009}
{Eyre} A.,  {Binney} J.,  2009, \mn@doi [\mnras]
  {10.1111/j.1365-2966.2009.15494.x}, \href
  {http://adsabs.harvard.edu/abs/2009MNRAS.400..548E} {400, 548}

\bibitem[\protect\citeauthoryear{{Eyre} \& {Binney}}{{Eyre} \&
  {Binney}}{2011}]{EyreBinney2011}
{Eyre} A.,  {Binney} J.,  2011, \mn@doi [\mnras]
  {10.1111/j.1365-2966.2011.18270.x}, \href
  {http://adsabs.harvard.edu/abs/2011MNRAS.413.1852E} {413, 1852}

\bibitem[\protect\citeauthoryear{{Gaia Collaboration} et~al.,}{{Gaia
  Collaboration} et~al.}{2016}]{GaiaDR12016}
{Gaia Collaboration} et~al., 2016, \mn@doi [\aap]
  {10.1051/0004-6361/201629512}, \href
  {http://adsabs.harvard.edu/abs/2016A%26A...595A...2G} {595, A2}

\bibitem[\protect\citeauthoryear{Grillmair}{Grillmair}{2016}]{Grillmair:2016ju}
Grillmair C.~J.,  2016, {Tidal Streams in the Local Group and Beyond}.
 Astrophysics and Space Science Library Vol. 420, Springer International
  Publishing, Cham

\bibitem[\protect\citeauthoryear{{Grillmair} \& {Carlin}}{{Grillmair} \&
  {Carlin}}{2016}]{Grillmair_book_2016}
{Grillmair} C.~J.,  {Carlin} J.~L.,  2016, in {Newberg} H.~J.,  {Carlin} J.~L.,
   eds,  Astrophysics and Space Science Library Vol. 420, Tidal Streams in the
  Local Group and Beyond. p.~87 (\mn@eprint {arXiv} {1603.08936}),
  \mn@doi{10.1007/978-3-319-19336-6_4}

\bibitem[\protect\citeauthoryear{{Grillmair} \& {Dionatos}}{{Grillmair} \&
  {Dionatos}}{2006}]{Grillmair2006}
{Grillmair} C.~J.,  {Dionatos} O.,  2006, \mn@doi [\apjl] {10.1086/505111},
  \href {http://adsabs.harvard.edu/abs/2006ApJ...643L..17G} {643, L17}

\bibitem[\protect\citeauthoryear{{Harris}}{{Harris}}{1976}]{Harris1976}
{Harris} W.~E.,  1976, \mn@doi [\aj] {10.1086/111991}, \href
  {http://adsabs.harvard.edu/abs/1976AJ.....81.1095H} {81, 1095}

\bibitem[\protect\citeauthoryear{{Ibata}, {Lewis}, {Irwin}  \& {Quinn}}{{Ibata}
  et~al.}{2002a}]{Ibata2002DM_TS}
{Ibata} R.~A.,  {Lewis} G.~F.,  {Irwin} M.~J.,   {Quinn} T.,  2002a, \mn@doi
  [\mnras] {10.1046/j.1365-8711.2002.05358.x}, \href
  {http://adsabs.harvard.edu/abs/2002MNRAS.332..915I} {332, 915}

\bibitem[\protect\citeauthoryear{{Ibata}, {Lewis}, {Irwin}  \&
  {Cambr{\'e}sy}}{{Ibata} et~al.}{2002b}]{Ibata2002PoleCount}
{Ibata} R.~A.,  {Lewis} G.~F.,  {Irwin} M.~J.,   {Cambr{\'e}sy} L.,  2002b,
  \mn@doi [\mnras] {10.1046/j.1365-8711.2002.05360.x}, \href
  {http://adsabs.harvard.edu/abs/2002MNRAS.332..921I} {332, 921}

\bibitem[\protect\citeauthoryear{{Ibata} et~al.,}{{Ibata}
  et~al.}{2017}]{CFIS_I_2017}
{Ibata} R.~A.,  et~al., 2017, \mn@doi [\apj] {10.3847/1538-4357/aa855c}, \href
  {http://adsabs.harvard.edu/abs/2017ApJ...848..128I} {848, 128}

\bibitem[\protect\citeauthoryear{{Johnston}, {Hernquist}  \&
  {Bolte}}{{Johnston} et~al.}{1996}]{Johnston1996}
{Johnston} K.~V.,  {Hernquist} L.,   {Bolte} M.,  1996, \mn@doi [\apj]
  {10.1086/177418}, \href {http://adsabs.harvard.edu/abs/1996ApJ...465..278J}
  {465, 278}

\bibitem[\protect\citeauthoryear{{Johnston}, {Spergel}  \& {Haydn}}{{Johnston}
  et~al.}{2002}]{Johnston2002DM_TS}
{Johnston} K.~V.,  {Spergel} D.~N.,   {Haydn} C.,  2002, \mn@doi [\apj]
  {10.1086/339791}, \href {http://adsabs.harvard.edu/abs/2002ApJ...570..656J}
  {570, 656}

\bibitem[\protect\citeauthoryear{{Kaiser} et~al.,}{{Kaiser}
  et~al.}{2002}]{PanSTARRS_Kaiser2002}
{Kaiser} N.,  et~al., 2002, in {Tyson} J.~A.,  {Wolff} S.,  eds,  \procspie
  Vol. 4836, Survey and Other Telescope Technologies and Discoveries. pp
  154--164, \mn@doi{10.1117/12.457365}

\bibitem[\protect\citeauthoryear{{King}}{{King}}{1966}]{King1966}
{King} I.~R.,  1966, \mn@doi [\aj] {10.1086/109857}, \href
  {http://adsabs.harvard.edu/abs/1966AJ.....71...64K} {71, 64}

\bibitem[\protect\citeauthoryear{{Koposov}, {Rix}  \& {Hogg}}{{Koposov}
  et~al.}{2010}]{Koposov2010}
{Koposov} S.~E.,  {Rix} H.-W.,   {Hogg} D.~W.,  2010, \mn@doi [\apj]
  {10.1088/0004-637X/712/1/260}, \href
  {http://adsabs.harvard.edu/abs/2010ApJ...712..260K} {712, 260}

\bibitem[\protect\citeauthoryear{{K{\"u}pper}, {Balbinot}, {Bonaca},
  {Johnston}, {Hogg}, {Kroupa}  \& {Santiago}}{{K{\"u}pper}
  et~al.}{2015}]{Kupper2015}
{K{\"u}pper} A.~H.~W.,  {Balbinot} E.,  {Bonaca} A.,  {Johnston} K.~V.,  {Hogg}
  D.~W.,  {Kroupa} P.,   {Santiago} B.~X.,  2015, \mn@doi [\apj]
  {10.1088/0004-637X/803/2/80}, \href
  {http://adsabs.harvard.edu/abs/2015ApJ...803...80K} {803, 80}

\bibitem[\protect\citeauthoryear{{Law} \& {Majewski}}{{Law} \&
  {Majewski}}{2010}]{LawMajewski2010}
{Law} D.~R.,  {Majewski} S.~R.,  2010, \mn@doi [\apj]
  {10.1088/0004-637X/714/1/229}, \href
  {http://adsabs.harvard.edu/abs/2010ApJ...714..229L} {714, 229}

\bibitem[\protect\citeauthoryear{{Marigo}, {Girardi}, {Bressan}, {Groenewegen},
  {Silva}  \& {Granato}}{{Marigo} et~al.}{2008}]{Marigo2008Padova}
{Marigo} P.,  {Girardi} L.,  {Bressan} A.,  {Groenewegen} M.~A.~T.,  {Silva}
  L.,   {Granato} G.~L.,  2008, \mn@doi [\aap] {10.1051/0004-6361:20078467},
  \href {http://adsabs.harvard.edu/abs/2008A%26A...482..883M} {482, 883}

\bibitem[\protect\citeauthoryear{{Mateu}, {Read}  \& {Kawata}}{{Mateu}
  et~al.}{2017}]{Mateu2017}
{Mateu} C.,  {Read} J.~I.,   {Kawata} D.,  2017, preprint, \href
  {http://adsabs.harvard.edu/abs/2017arXiv171103967M} {} (\mn@eprint {arXiv}
  {1711.03967})

\bibitem[\protect\citeauthoryear{{Myeong}, {Jerjen}, {Mackey}  \& {Da
  Costa}}{{Myeong} et~al.}{2017}]{Eridanus_pal15_2017}
{Myeong} G.~C.,  {Jerjen} H.,  {Mackey} D.,   {Da Costa} G.~S.,  2017,
  preprint, \href {http://adsabs.harvard.edu/abs/2017arXiv170407690M} {}
  (\mn@eprint {arXiv} {1704.07690})

\bibitem[\protect\citeauthoryear{{Ngan} \& {Carlberg}}{{Ngan} \&
  {Carlberg}}{2014}]{Ngan2014}
{Ngan} W.~H.~W.,  {Carlberg} R.~G.,  2014, \mn@doi [\apj]
  {10.1088/0004-637X/788/2/181}, \href
  {http://adsabs.harvard.edu/abs/2014ApJ...788..181N} {788, 181}

\bibitem[\protect\citeauthoryear{{Odenkirchen} et~al.,}{{Odenkirchen}
  et~al.}{2001}]{Odenkirchen2001}
{Odenkirchen} M.,  et~al., 2001, \mn@doi [\apjl] {10.1086/319095}, \href
  {http://adsabs.harvard.edu/abs/2001ApJ...548L.165O} {548, L165}

\bibitem[\protect\citeauthoryear{{Robin} et~al.,}{{Robin}
  et~al.}{2012}]{GUMS2012}
{Robin} A.~C.,  et~al., 2012, \mn@doi [\aap] {10.1051/0004-6361/201118646},
  \href {http://adsabs.harvard.edu/abs/2012A%26A...543A.100R} {543, A100}

\bibitem[\protect\citeauthoryear{{Rockosi} et~al.,}{{Rockosi}
  et~al.}{2002}]{Rockosi2002}
{Rockosi} C.~M.,  et~al., 2002, \mn@doi [\aj] {10.1086/340957}, \href
  {http://adsabs.harvard.edu/abs/2002AJ....124..349R} {124, 349}

\bibitem[\protect\citeauthoryear{{Sanders}, {Bovy}  \& {Erkal}}{{Sanders}
  et~al.}{2016}]{StreamGap_Sanders2016}
{Sanders} J.~L.,  {Bovy} J.,   {Erkal} D.,  2016, \mn@doi [\mnras]
  {10.1093/mnras/stw232}, \href
  {http://adsabs.harvard.edu/abs/2016MNRAS.457.3817S} {457, 3817}

\bibitem[\protect\citeauthoryear{{Sch{\"o}nrich}, {Binney}  \&
  {Dehnen}}{{Sch{\"o}nrich} et~al.}{2010}]{Schornich2010_Sun}
{Sch{\"o}nrich} R.,  {Binney} J.,   {Dehnen} W.,  2010, \mn@doi [\mnras]
  {10.1111/j.1365-2966.2010.16253.x}, \href
  {http://adsabs.harvard.edu/abs/2010MNRAS.403.1829S} {403, 1829}

\bibitem[\protect\citeauthoryear{{Shipp} et~al.,}{{Shipp}
  et~al.}{2018}]{DES_11_streams2018}
{Shipp} N.,  et~al., 2018, preprint, \href
  {http://adsabs.harvard.edu/abs/2018arXiv180103097S} {} (\mn@eprint {arXiv}
  {1801.03097})

\bibitem[\protect\citeauthoryear{{Smith} et~al.,}{{Smith}
  et~al.}{2007}]{EscapeVel2007}
{Smith} M.~C.,  et~al., 2007, \mn@doi [\mnras]
  {10.1111/j.1365-2966.2007.11964.x}, \href
  {http://adsabs.harvard.edu/abs/2007MNRAS.379..755S} {379, 755}

\bibitem[\protect\citeauthoryear{{Teuben}}{{Teuben}}{1995}]{Teuben1995NEMO}
{Teuben} P.,  1995, in {Shaw} R.~A.,  {Payne} H.~E.,   {Hayes} J.~J.~E.,  eds,
  Astronomical Society of the Pacific Conference Series Vol. 77, Astronomical
  Data Analysis Software and Systems IV. p.~398

\bibitem[\protect\citeauthoryear{{Thomas}, {Ibata}, {Famaey}, {Martin}  \&
  {Lewis}}{{Thomas} et~al.}{2016}]{Thomas2016}
{Thomas} G.~F.,  {Ibata} R.,  {Famaey} B.,  {Martin} N.~F.,   {Lewis} G.~F.,
  2016, \mn@doi [\mnras] {10.1093/mnras/stw1189}, \href
  {http://adsabs.harvard.edu/abs/2016MNRAS.460.2711T} {460, 2711}

\bibitem[\protect\citeauthoryear{{Varghese}, {Ibata}  \& {Lewis}}{{Varghese}
  et~al.}{2011}]{Varghese2011}
{Varghese} A.,  {Ibata} R.,   {Lewis} G.~F.,  2011, \mn@doi [\mnras]
  {10.1111/j.1365-2966.2011.19097.x}, \href
  {http://adsabs.harvard.edu/abs/2011MNRAS.417..198V} {417, 198}

\bibitem[\protect\citeauthoryear{{Williams} et~al.,}{{Williams}
  et~al.}{2011}]{Aquarius_Williams_2011ApJ...728..102W}
{Williams} M.~E.~K.,  et~al., 2011, \mn@doi [\apj]
  {10.1088/0004-637X/728/2/102}, \href
  {http://adsabs.harvard.edu/abs/2011ApJ...728..102W} {728, 102}

\bibitem[\protect\citeauthoryear{{de Bruijne}}{{de
  Bruijne}}{2012}]{Gaia2012deBruijne}
{de Bruijne} J.~H.~J.,  2012, \mn@doi [\apss] {10.1007/s10509-012-1019-4},
  \href {http://adsabs.harvard.edu/abs/2012Ap%26SS.341...31D} {341, 31}

\makeatother
\end{thebibliography}


\bsp	
\label{lastpage}
\end{document}